\documentclass[12pt,a4paper]{article}

\usepackage[utf8]{inputenc}
\usepackage[T1]{fontenc}
\usepackage[margin=2.5cm]{geometry}
\usepackage[numbered]{bookmark}
\usepackage{mathtools}
\usepackage{amssymb}
\usepackage{graphicx}
\usepackage[font=small,labelfont=bf]{caption}
\usepackage{authblk}
\usepackage{cite}
\usepackage{dsfont}
\usepackage{xcolor}
\usepackage[titles]{tocloft}

\usepackage[displaymath, mathlines]{lineno} 

\numberwithin{equation}{section}

\newcommand{\eq}[1]{\begin{equation} #1 \end{equation}}
\newcommand{\al}[1]{\begin{align} #1 \end{align}}

\newcommand{\pa}{\partial}

\newcommand{\sn}{\mathrm{sn}}

\title{On Pulsating Strings in Schr\"{o}dinger Backgrounds}

\author[a,b]{H.~Dimov}
\author[a]{M.~Radomirov} 
\author[a,c]{R.~C.~Rashkov}
\author[a,b]{T.~Vetsov\footnote{Emails: \texttt{h\_dimov,rash,vetsov@phys.uni-sofia.bg} and \texttt{miroslavsky@abv.bg}}
}

\affil[a]{\textit{Department of Physics, Sofia University,}\authorcr\textit{5 J. Bourchier Blvd., 1164 Sofia, Bulgaria}\vspace{5pt} \vspace{3pt}} 
\affil[b]{\textit{The Bogoliubov Laboratory of Theoretical Physics, JINR,}\authorcr\textit{141980 Dubna, Moscow region, Russia}\vspace{5pt} \vspace{3pt}} 
\affil[c]{\textit{Institute for Theoretical Physics, Vienna University of Technology,}\authorcr\textit{Wiedner Hauptstr. 8--10, 1040 Vienna, Austria}}

\date{}
	
\begin{document}
	 
\maketitle
\begin{abstract}
According to AdS/CFT duality semi-classical strings in the Schr\"odinger spacetime is conjectured to be a holographic dual to dipole CFT.
In this paper we consider pulsating strings in five-dimensional Schr\"odinger space times five-sphere. We have found classical string solutions pulsating entirely in the Schr\"odinger part of the background. We quantize the theory semi-classically and obtain the wave function of the problem. We have found the corrections to the energy, which by duality are supposed to give anomalous dimensions of certain operators in the dipole CFT. 	
\end{abstract}

\section{Introduction}

In the last two decades the celebrated AdS/CFT correspondence has proven to be a powerful tool for studying important aspects of string theory and conformal field theories. First thing to take care of is to establish a dictionary between objects on both sides of the duality.
Started with examples of theories with high amount of supersymmetry in supergravity approximation, the correspondence has been developed beyond this approximation, namely taking into account essentially string modes. 
Although solving exactly the string spectrum on a generic background is highly non-trivial problem, the discovery of rich set of integrable structures within the correspondence allowed the fruitful study of many string configurations.
In this context, a large variety of rotating strings, spinning strings, giant magnons, folded strings, spiky strings, pulsating strings and their gauge theory duals have been analyzed in great details.

Addressing the important issues as strong coupling phenomena, it became of great interest to extend it to less supersymmetric theories, which are phenomenologically more appropriate. From gravitational point of view, breaking supersymmetry means deformation of the geometry, such that  the new background is again an Einstein manifold. 
The remarkably simple procedure advocated in  \cite{Lunin:2005jy} enables to explicitly generate new solutions as well as to suggest what holographic duals of these solutions could be. This approach is based on the global symmetries underlying undeformed theory. Concretely, having two global $U(1)$ isometries one can interpret them geometrically as a two-torus. The action of the associated $SL(2,\mathbb{R})$ symmetry on the torus parameter is  $\tau\to \tilde{\tau}=\tau/(1+\gamma\tau)$. Specializing to type  IIB backgrounds, it is known that we already have $SL(2,\mathbb{R})$ symmetry of ten dimensional background. The two $SL(2,\mathbb{R})$ symmetries get combined to form $SL(3,\mathbb{R})$ (which can be seen compactifying M-theory on $T^3$). The explicit form of the $SL(3,\mathbb{R})$ matrix used to generate non-singular solutions is
\eq{
g_3=\begin{pmatrix}1 & 0 & 0\\ \gamma & a & b\\ \sigma & c & d \end{pmatrix}\in SL(3,\mathbb{R}), \qquad g_2=\begin{pmatrix} a & b \\ c & d\end{pmatrix}\in SL(2,\mathbb{R}).
}

From holographic point of view however it is important where in the space-time the two-torus associated with the global isometries lives.
Indeed, if the torus lives entirely in the (asymptotically) AdS part of the background geometry the dual gauge theory becomes non-commutative. The $U(1)$ charges are actually momenta and the product in the field theory is replaced by star-product. If however the torus is in the complementary part, in the field theory the product is the ordinary product, but the theory exhibits a Leigh-Strassler deformation \cite{Leigh:1995ep}.  Since the deformation uses  a direction along the R-symmetry it is quite clear that the supersymmetry in the new theory will be at least partially broken. This is in agreement with the gauge theory side \cite{Leigh:1995ep}.

Gauge theories with deformed products of fields and their string theory duals, obtained
with or without the knowledge of the $SL(2,R)$ transformation, have been thoroughly studied
in the literature. Some recent advancements on the study of Lunin-Maldacena background can be found for instance in \cite{Frolov:2005ty,frolovnew, Gursoy:2005cn, Chu:2006ae, Bobev:2006fg, Bobev:2007bm, Bobev:2005cz, Bykov:2008bj, Dimov:2009ut, Frolov:2005dj}.

The solution generating technique can be applied to wide range of cases.
The most well-known example is probably the one of exactly marginal deformations of (super) conformal gauge theories, such as the deformations of $\mathcal N = 4$ Super Yang-Mills theory, considered in \cite{Leigh:1995ep}, whose gravity dual (for the so-called $\beta$-deformation) was derived in \cite{Lunin:2005jy}.
 
Another important case leading to qualitatively different solutions is the one dual to ''dipole'' theories \cite{BG, IKK, oj}. It is characterized with
non-locality of the dual gauge theory but still living on an ordinary commutative space. One direction, where these solutions have been used, is in description of ordinary $\mathcal N = 1$ SQCD-like gauge theories in the context of D-brane realizations. This type of deformations are typically used  to decouple  spurious effects coming from Kaluza-Klein modes on cycles  wrapped by the D-brane \cite{Gursoy:2005cn, Chu:2006ae, Bobev:2006fg, Bobev:2007bm,Bobev:2005cz}.

Recently there has been revival of interest to holography in Schr\"odinger backgrounds. It has been inspired by the attempts to generalize the AdS/CFT correspondence to strongly coupled non-relativistic conformal theories \cite{Son:2008ye,Balasubramanian:2008dm}. In these cases the isometry group of the background geometry on string side is the Schr\"odinger group. It consists of time and space translations, space rotations, Galilean boosts and dilatations.
It has been shown that one possible dual to such theory  could be dipole theories. Further progress of this line of investigations can be traced for instance in \cite{BG,IKK,oj}. An important step has been done in \cite{Guica:2017jmq} where strong arguments for integrability and quantitative check of the matching between string and gauge theory predictions have been presented. This triggered a new interest to holography in such backgrounds, since this opens the option to get important information for dipole theories for instance. Particular string solutions beyond the supergravity approximation has been found in \cite{Guica:2017jmq,Ahn:2017bio,Georgiou:2017pvi,Georgiou:2018zkt}. A semiclassical quantization has been also considered in \cite{Ouyang:2017yko}.

In this paper we focus on the duality with non-relativistic gauge theory dual. This means that we have to consider background respecting Schr\"odinger symmetry. On the gauge theory side the theories are known as non-relativistic duals. To obtain the Schr\"odin-ger background we will use TsT deformations, which involve the time direction and one spatial dimension in the internal space. As it was shown for instance in \cite{Bobev:2009mw}, the generated solutions are twisted and the supersymmetry becomes completely broken. We will describe this technique in some details in the next Section, where we will also briefly review pulsating strings in holography. In the paper we restrict our considerations to the bosonic sector of the theory.


\section{How to generate Schr\"odinger backgrounds}

We first begin with an overview of the methods of deformations used later to generate  Schr\"odinger spaces.

The purpose of the  solution-generating technique suggested in \cite{Lunin:2005jy} is to obtain the gravity duals of the deformed gauge theories or such with reduced (super)symmetry. Starting from a supergravity solution of a type II theory, applying a simple procedure called TsT transformation, one obtains another supergravity solution. The procedure of TsT transformation is described as follows. First of all, one has to identify two isometry directions which can be thought of as two-torus. Let us parametrize the two torus angles by $(\phi_1,\phi_2)$. The transformation then consists of a T-duality along $\phi_1$, followed by a shift $\phi_2 \rightarrow \phi_2 + \gamma\phi_1$ in the T-dual background. The final step is to T-dualize back along $\phi_1$.

This simple but remarkably effective transformations can be readily applied to a number of backgrounds, having these properties and particular, in the contexts of string/gauge theory duality. 

Depending on where the isometry directions involved in TsT transformations live, we distinguish three cases of deformations:
\begin{enumerate}
	\item To implement the first class of deformations, we take the two isometries involved in the TsT-transformation to be along the brane. In this case the product of the fields in the dual gauge theory becomes:
	\begin{eqnarray}
	(f\ast g)(x)& =& e^{-i\pi\gamma\left(
		\frac{\partial}{\partial x^1}\frac{\partial}{\partial y^2}-\frac{\partial}{\partial x^2}
		\frac{\partial}{\partial y^1}\right)}
	f(x)g(y)_{|x=y}\nonumber \\
	& = & f(x)g(x) - i\pi\gamma\left(\partial_{1}f(x)\partial_2g(x)-
	\partial_2f(x)\partial_1g(x)\right)+\cdots
	\end{eqnarray}
Since gamma is constant, one can easily recognize this product as the Moyal product for a non-commutative two-torus. One should note also that this deformation makes the theory non-commutative. Obviously it is non-local and breaks Lorentz invariance and causality. Nevertheless, this picture, and its generalizations, opens the road to interesting string realizations of non-commutative theories. The generalization to more non-commutative directions is straightforward.
	
	\item  To make the second class of deformations we assume that there is one global $U(1)$ isometry along the D-brane, but the other one, which is going to be used for TsT-transformation, is transversal to the brane. 	If we associate charges to the isometries, say $Q^i$, one can map the string picture to the gauge theory dual.  In this case one of the charges remains to be the momentum but the other one is not anymore.  The deformed product of the fields in this case can be read off:
\begin{equation}
(f\ast g)(x)=e^{\pi\gamma\left(
Q^g\frac{\partial}{\partial x}-Q^f\frac{\partial}{\partial y^1}\right)}=
f(x+\pi\gamma Q^g)g(x-\pi\gamma Q^f).
\end{equation}
This is called dipole deformation. As we can see it is clearly non-local in one direction, but still living on a commutative space-time. Here, we applied TsT-transformation shifting
single direction $x$, but we can obtain more general deformations by introducing ''dipole vectors'', $L^{M\mu} =-2\pi\gamma Q^ML^\mu$, for the various fields (here $L^\mu$ is a constant vector). In this case, the product of the fields is
\begin{equation}
(f\ast g)(x)=f(x-\frac{1}{2}L^g)g(x+\frac{1}{2}L^f).
\label{star-dipole}
\end{equation}
	\item  Finally, the last case involves two isometries transverse to the D-brane. The dual gauge theory picture changes since the two charges does not act as derivatives. Thus, the product of the fields reduces to
\begin{equation}
(f\ast g)(x)=e^{i\pi\gamma(Q^f_1Q^g_2-Q^f_2Q^g_1)}fg. \label{2.6}
\end{equation}
In this case, the product deformation yields an ordinary commutative and local theory,
since the only contribution to the deformed product \eqref{2.6}  is to introduce some
phases in the interactions. The superconformal gauge theories, arising from the product \eqref{2.6}, are classified by Leigh and Strassler and are called $\beta$-deformed.
One can note that the deformation reduces the amount of supersymmetry and TsT-transformations can serve as supersymmetry breaking procedure.
\end{enumerate}

The most general deformation, which contains the above-mentioned techniques is
based on the so-called Drinfe’ld-Reshetikhin (DR) twist. For completeness, in an Appendix \ref{appA} we give a
very brief review of these deformations following mainly \cite{Guica:2017jmq}.

It is well known that the dipole deformations have as a dual non-relativistic theory, see for instance \cite{BG,IKK,oj}.
In general, the deformations, related to non-relativistic theories, are a little bit specific and we collect here in a systematic way the known facts from the literature.
The procedure of the solution generating technique, we will follow below, consists in the following steps:
\begin{enumerate}
	\item\label{step1} The starting point is a solution of type IIA/IIB string background and the first step is to recognize the compact/non-compact translationally invariant directions. Let us denote the Killing vector of one of them by $(\pa/\pa y)^a$.
	\item\label{step2} Boost the geometry in $y$ direction by an amount $\gamma$. It effectively is charging the initial solution by a momentum charge. However, if we start with an $SO(1,1)$ symmetry, one can pair the Killing vector $(\pa/\pa y)^a$ with the timelike one $(\pa/\pa t)^a$ and thus, no additional charge is added. Then, this step appears to be trivial, i.e. diagonal, since the theory is boost-invariant.
	\item\label{step3}  Perform a T-duality along $y$. This brings us to IIB/IIA supergravity solution for the background.
	\item\label{step4} Generation of the twist can be conducted considering some additional $S^1$ or translational isometries. Let the one-form, associated with the additional isometry, is $\sigma$. Then, we perform a twist by an amount of $\alpha$
\eq{
		\sigma\:\rightarrow\: \sigma + 2\alpha dy.
	}
	\item\label{step5} We now T-dualize the geometry back to IIA/IIB along $y$. The twist followed by the
	T-duality is effectively a non-diagonal T-duality.
	\item\label{step6} Boost along $y$ by $-\gamma$ to undo the initial boost.
	\item\label{step7} For the decoupling we make a double-scaling limit keeping finite the ratio $\alpha e^\gamma$:
	\eq{
		\beta=\frac{1}{2}\alpha e^\gamma = \operatorname{fixed}.
	}
\end{enumerate}
The procedure, outlined above, is known also as null Melvin twist transformations. The steps \ref{step4} through \ref{step7} can be thought of as converting
the string solution into a fluxbrane, followed by a boost and scaling to end up with a null
isometry.

\indent In addition to the discussion above, a large class of deformations can be found in \cite{Bobev:2009mw,Bobev:2009zf}. A more general conclusion from there can be drown. Namely, the general form of large class of background solutions, obtained by null Melvin twists, have the form:
\al{
	& ds^2_{10}=-\frac{\Omega}{z^{2n}}du^2 + \frac{1}{z^2} (-2dvdu + dx_1^2+ dx_2^2 + dz^2) + ds^2_{X^5},\\
	& F_{(5)}=(1+\star)\text{vol}_{X^5}, \\
	& B_{(2)}=\frac{1}{z^{2n}}P\wedge du,
}
where $X^5$ is an Einstein manifold and $P$ is one-form on $X^5$. In addition to the vector eigenfunction of the Laplacian on $X^5$, the generic background of this type includes also a B-field. Moreover,  the factor $\Omega$ in the metric obeys an inhomogeneous scalar Laplace equation on $X^5$. We note that the analysis conducted in this paper for the particular case of $Schr_5\times S^5$ background can be easily extended to these cases.

Actually, for $S^5$, this is the Hopf fibration $S^1\rightarrow S^5\rightarrow\mathbb{CP}^2$. Thus, $P$ is the 1-form potential for the K\"ahler form on $\mathbb{CP}^2$, and therefore $J_P^2 = dP$.

Integrability, which plays a crucial role in proving holography, has been considered in some details in \cite{Guica:2017jmq}. 
Concretely, consider $AdS_5\times X^5$, where the metric on $X^5$ is $g_{\alpha\beta}$. Now we perform a null Melvin twist along a Killing vector $\mathcal{K}$ on $X^5$. The result is
\eq{
	ds^2_{10}=ds^2_{Schr_5}+ds^2_{X^5},
}
where the Schr\"odinger part is given by
\eq{
	ds^2_{Schr_5}=-\frac{\Omega}{z^4}+\frac{1}{z^2}(-2dvdu+dx_1^2+dx_2^2+dx_3^2), \quad \Omega= ||\mathcal{K}||^2=g_{\alpha\beta}\mathcal{K}^\alpha \mathcal{K}^\beta.
}
It is clear that $\Omega$ is non-negative being a square length of the Killing vector. The generated $B$-field has the form ($\mathcal{K}_\alpha=g_{\alpha\beta} \mathcal{K}^\beta$):
\eq{
	B_{(2)}=\frac{1}{z^2}\mathcal{K}\wedge du.
}

An important remark is that in order
to make holographic sense of these solutions, one has to impose some conditions. In order these to be holographic duals to non-relativistic field theories the light-cone coordinate $v$ should be periodic, $v\sim v +2\pi r_v$ \cite{Son:2008ye,Balasubramanian:2008dm,Adams:2008wt}. The momentum along this compact direction is quantized in units of the inverse radius $r_v^{-1}$. This momentum is interpreted as the Galilean mass (or the particle number) in the dual field theory. Note that the norm of the Killing vector $\mathcal{K}$ may vanish on some locus in $X^5$, but still having a perfect gravity dual of non-relativistic theory\footnote{This point is discussed, for instance in \cite{Hubeny:2005qu,Gimon:2003xk}.}!
In \cite{Guica:2017jmq} the integrability has been discussed in the context of Bethe ansatz and certain properties, like spinning BNM solutions and their dispersion relations have been analyzed.

For particular examples, further discussions and procedures see also \cite{Hubeny:2005qu,Gimon:2003xk}.

\paragraph{The case of $AdS_5\times S^5$}\ \\

\noindent Let us write the metric of the space-time as
\eq{
	ds^2=ds^2_{AdS_5}+ds^2_{S^5}.
}
Here, the metric on $AdS_5$, in light-cone coordinates, is written by
\eq{
	ds^2_{AdS_5}=\ell^2\,\frac{2dx^+dx^-+dx^idx_i +dz^2}{z^2},
	\label{ads-l-c}
}
where $\ell$ is the AdS radius in the string frame. The metric of $S^5$ is useful to write as
\begin{equation}
ds^2_{S^5}=ds_{\mathbb{CP}^2}^2+(e^5)^2.
\end{equation}
The two terms are the metric of the round $S^5$ in the form of Hopf fibration.
We introduce the coordinate $\chi$ for the Hopf fibers, and use the local orthonormal frame
\begin{equation}
e^m\quad(m=1,2,3,4),\quad
e^5=d\chi+P.
\end{equation}
Here $e^m$ are the vielbein in the base $\mathbb{CP}^2$ and $P$ is a differential on it
\begin{equation}
P=\frac{1}{2}\sin^2\mu\left(d\alpha+\cos\theta\,d\phi\right).
\end{equation}
Its exterior derivative is proportional to the K\"ahler form $I_{\mathbb{CP}^2}$ on $\mathbb{CP}^2$,
\begin{equation}
dP=-\frac{2}{r}I_{\mathbb{CP}^2}
=-\frac{1}{r}I_{mn}e^m\wedge e^n=\frac{2}{r}(e^1\wedge e^2+e^3\wedge e^4).
\end{equation}
The final form for the metric becomes
\eq{
		ds^2=\ell^2\left(\frac{2dx^+dx^-+dx^idx_i+dz^2}{z^2}+ (d\chi+P)^2+ds^2_{\mathbb{CP}^2}  \right).
}
\indent The program we are going to follow is:
\begin{itemize}
	\item Make a T-duality along $\chi$.
	\item Make a shift $x^-\,\rightarrow\, x^-+\tilde{\mu}\tilde{\chi}$, where $\tilde{\chi}$ is T-dualized $\chi$.
	\item Make T-duality back on $\tilde{\chi}$.
\end{itemize}
Following these rules one easily finds
\eq{
		ds^2=\ell^2\left(- \frac{{\hat\mu}^2(dx^+)^2}{z^4} + \frac{2dx^+d\hat{x}^-+dx^idx_i+dz^2}{z^2} \right)+ ds^2_{\hat{S}^5} ,
	\label{schro-metric}
}
where the metric on the five-sphere is
\begin{align} \label{eq_S5_metric}
\frac{ds^2_{\hat S^5}}{\ell^2}&=\,d\hat\chi^2+d\mu^2+\frac{1}{4}\sin^2\mu\left(d\alpha^2+d\theta^2+d\phi^2\right)\nonumber\\
&+\sin^2\mu\,d\hat\chi\,d\alpha+\sin^2\mu\cos\theta\,d\hat\chi\,d\phi+\frac{\sin^2\mu\cos\theta}{2}\,d\alpha\,d\phi.
\end{align}
Furthermore, the deformed TsT background acquires non-zero $B$-field:\footnote{Note that in the original $AdS_5\times S^5$ theory there is no $B$-field.}
\eq{
	\alpha' B_{(2)}=\frac{\ell^2{\hat\mu}\, dx^+}{z^2}\wedge (d\hat{\chi}+ P).
	\label{schro-B}
}
To complete the discussion, we write the metric of the Schr\"odinger background and the B-field in global coordinates (for details see for instance \cite{Blau:2009gd})
\eq{
\frac{ds^2_{Schr_5}}{\ell^2}=-\left(\frac{{\hat\mu}^2}{Z^4}+1 \right)dT^2+
\frac{2dT\,dV-\vec{X}^2dT^2+d\vec{X}^2+dZ^2}{Z^2},
	\label{metric-schro-global}
}
\eq{
\alpha' B_{(2)}= \frac{\ell^2{\hat\mu}\, dT}{Z^2}\wedge (d\hat{\chi}+P), \qquad
\hat\mu=\frac{\ell^2}{\alpha'}\tilde{\mu}=\sqrt{\lambda}\tilde{\mu}=\frac{\sqrt{\lambda}}{2\pi}L.
	\label{B-global-12}
}
Here $\hat{x}^-$ and $\hat{\chi}$ are the dualized coordinates, $L$ is the shift parameter in the star product in \eqref{star-dipole}. The relation between the original and dualized coordinates is
\al{
 d\chi= d\hat{\chi} + \hat{\mu}\frac{dx^+}{z^2} \label{dual-psi},\qquad dx^-=d\hat{x}^- -\hat{\mu}\left(d\hat{\chi} +\hat{\mu} \frac{dx^+}{z^2} +P\right).
}
\begin{itemize}
	\item 
	The dual coordinates satisfy periodic boundary conditions.
	\item 
	The original coordinates satisfy twisted boundary conditions:
	\end{itemize}
\begin{equation}
\label{eq_boundary_condition_1}
		x^-(2\pi)-x^-(0)= LJ,
\end{equation}
\begin{equation}\label{eq_boundary_condition_2}
		\chi(2\pi)-\chi(0)=2\pi m - LP_-,
\end{equation}

where $m\in \mathbb{Z}$ and $P_-$ is the charge associated with the null Melvin twist.

In the case of Schr\"odinger space times five sphere the twisted solution generated by TsT-transformations break supersymmetry completely \cite{Bobev:2009mw}. In what follows we will focus on the bosonic sector of the theory.

\section{Pulsating strings in Schr\"odinger background}

We start this section with an overview of the pulsating strings in the most supersymmetric example of AdS/CFT correspondence, namely $AdS_5\times S^5$. Next we apply this construction to obtain pulsating string solutions in Schr\"odinger background.

\subsection{Pulsating strings in AdS/CFT correspondence}

Let us briefly discuss pulsating strings in $AdS_5\times S^5$ space. Pulsating strings were first introduced in \cite{min} 
and later on were generalized by \cite{Engquist:2003, Dimov:2004, Smedback:2004}. Since then there have been further examples of pulsating strings, both
in AdS and non-AdS backgrounds\cite{Khan:2003sm, 
	Arutyunov:2003za, Kruczenski:2004cn, Bobev:2004id, Park:2005kt, deVega:1994yz, Chen:2008qq, Dimov:2009rd, Arnaudov:2010by, Arnaudov:2010dk, Beccaria:2010zn, Giardino:2011jy, Pradhan:2013sja, Pradhan:2014zqa}. In this section we give a brief review on the pulsating string solution obtained in \cite{min}. 
\indent Let us consider a circular string, which pulsates on $S^5$ by expanding and contracting its length. In this case, the metric of $S^5$ and the relevant part of $AdS_5$ are given by
\eq{
	ds^2=R^2\left(\cos^2\theta d\Omega_3^2+d\theta^2+\sin^2\theta d\psi^2+
	d\rho^2-{\rm{cosh}}^2\rho dt^2\right),
	\label{1.1}
}
where $R^2={\alpha}^\prime\sqrt{\lambda}$ with $\lambda$ the ’t Hooft coupling. One can obtain the simplest pulsating string solution by identifying the target space time coordinate $t$ with the worldsheet
one, $t=\tau$, and setting $\psi=m\sigma$, which corresponds to a string stretched along $\psi$ direction. We also set the ansatz for $\theta=\theta(\tau)$ and  $\rho=\rho(\tau)$. Hence, the Nambu-Goto action reduces to
\eq{
	S=-m\sqrt{\lambda}\int dt \sin\theta \sqrt{{\rm{cosh}}^2\rho-\dot{\rho}^2 -\dot\theta^2}\,,
	\label{1.1a}
}
where $\dot{\theta}=d\theta/d\tau$. In order to obtain the solution and the string spectrum it is useful to pass to Hamiltonian
formulation. For this purpose, after identifying the canonical momenta,
\begin{equation}
{\Pi _\rho }\; = \;\frac{{m\sqrt \lambda  \;\sin \theta \;\dot \rho }}{{\sqrt {{{\cosh }^2}\rho  - {{\dot \rho }^2} - {{\dot \theta }^2}} }},\qquad {\Pi _\theta }\; = \;\frac{{m\sqrt \lambda  \;\sin \theta \;\dot \theta }}{{\sqrt {{{\cosh }^2}\rho  - {{\dot \rho }^2} - {{\dot \theta }^2}} }},
\end{equation}
we can write the Hamiltonian in the form \cite{min}:
\eq{
	H={\rm{cosh}}\rho\sqrt{\Pi_\rho^2+\Pi_\theta^2+m^2\lambda\sin^2\theta}.
	\label{1.2}
}
If the string is placed at the origin ($\rho=0$) of $AdS_5$ space, we
see that the squared Hamiltonian have a form similar to a point
particle. Here, the last term in (\ref{1.2}) can be considered as a
perturbation. Therefore one can first find the wave function for a free particle
in the above geometry
\begin{equation}\label{1.3}
- \frac{{\cosh \rho }}{{{{\sinh }^3}\rho }}\frac{d}{{d\rho }}\left( {\cosh \rho {{\sinh }^3}\rho \frac{d}{{d\rho }}\psi (\rho ,\theta )} \right) - \frac{{{{\cosh }^2}\rho }}{{\sin \theta {{\cos }^3}\theta }}\frac{d}{{d\theta }}\left( {\sin \theta {{\cos }^3}\theta \frac{d}{{d\theta }}\psi (\rho ,\theta )} \right) = {E^2}\psi (\rho ,\theta ).
\end{equation}
Standard separation of variables, $\psi (\rho ,\theta )=f(\rho)\, g(\theta)$, leads to two decoupled ordinary differential equations, namely an equation for $g(\theta)$
\begin{equation}\label{eqTheta}
\frac{{1 }}{{g(\theta)\sin \theta {{\cos }^3}\theta }}\frac{d}{{d\theta }}\left( {\sin \theta {{\cos }^3}\theta \frac{d}{{d\theta }}g (\theta )} \right)=\alpha,
\end{equation}
where $\alpha$ is the separation constant, and an equation for $f(\rho)$
\begin{equation}\label{eqRho}
\frac{{\cosh \rho }}{{{{\sinh }^2}\rho }}\frac{d}{{d(\cosh\rho )}}\left( {\cosh \rho {{\sinh }^4}\rho \frac{d}{{d(\cosh\rho )}}f (\rho  )} \right) + \alpha\,{{\cosh }^2}\rho\,f(\rho) + {E^2} f (\rho )=0.
\end{equation}
Equation (\ref{eqTheta}) reduces to 
\begin{equation}
g''(\theta ) + (\cot \theta  - 3\tan \theta )g'(\theta ) - \alpha g(\theta ) = 0.
\end{equation}
Its regular solution is proportional to a hypergeometric function,
\begin{equation}
g(\theta ) = {{\mkern 1mu} _2}{F_1}\left( {1 - \frac{{\sqrt {4 - \alpha } }}{2},\frac{1}{2}\left( {\sqrt {4 - \alpha }  + 2} \right),2,{{\cos }^2}\theta } \right),
\end{equation}
which reduces to a polynomial if its series is truncated at some finite integer order $n$. This can be achieved if we set its first argument to be equal to $-n$, thus we find the separation constant $\alpha$:
\begin{equation}\label{eqSeparAlpha}
\alpha  =  - 4n(n + 2).
\end{equation}
Taking advantage of the relation between the hypergeometric function and the Jacobi polynomials, namely
\begin{eqnarray}
_2{F_1}( - n,\tilde \alpha  + \tilde \beta  + 1 + n,\tilde \alpha  + 1,z) = \frac{{n!}}{{{{(\tilde \alpha  + 1)}_n}}}P_n^{(\tilde \alpha ,\tilde \beta )}(1 - 2z),
\end{eqnarray}
where, in our case $\tilde\alpha=1$, $\tilde\beta=0$ and $z=\cos\theta$. We can further transform the Jacobi polynomial to the standard spherical harmonics $P_{2n}(\cos\theta)$, i.e. $P_m^{(1 ,0 )}(1 - 2z)=c P_{m}(z)$ \cite{Braaksma_1968}, where $m=2n$ should be even. Thus the final polynomial solution of Eq. (\ref{eqTheta}) up to a normalization constant is
\begin{equation}
g(\theta ) = P_{2n}(\cos\theta).
\end{equation}

Let us take a look at the second equation (\ref{eqRho}). Changing the variable $\rho$ to $x=\cosh\rho >0 $, one finds
\begin{equation}
x^2\,(x^2-1)\, f''(x) +(5x^2-1)\,x\,f'(x) +\left( \alpha\,x^2 + E^2 \right) \, f(x) =0.
\end{equation}
We can look for simple polynomial solution of type $f(x)= c x^a$, where $a$ is some constant. This leads to
\begin{equation}
c{x^a}\left( {{x^2}\left( {{a^2} + 4a + \alpha } \right) - {a^2} + {E^2}} \right) = 0,
\end{equation}
which is satisfied only if
\begin{equation}
{a^2} + 4a =  - \alpha ,\quad {E^2} - {a^2} = 0.
\end{equation}
Substituting $\alpha$ from equation (\ref{eqSeparAlpha}) one finds $a=-2 n-4$ and the full solution to Eq. (\ref{1.3}) becomes
\eq{
	{\Psi _{2n}}(\rho ,\theta ) = {(\cosh \rho )^{ - 2n - 4}}{\mkern 1mu} P_{2n}(\cos\theta).
	\label{1.4}
}
Now, the energy levels are given by
\eq{
	E=\Delta=2n+4.
	\label{1.5}
}

As we mentioned in the Introduction, strings in Schr\"odinger background is conjectured to have as a holographic dual strongly coupled non-relativistic conformal theory. According to holographic dictionary the energy on string side should correspond to the (anomalous) dimension of certain operator on field theory side. Thus, \eqref{1.5} is interpreted as the bare dimension of the field theory operator. The weak coupling on string theory side allows to expand \eqref{1.2} in $\lambda$ and obtain the first quantum corrections.
For highly excited states (large energies), one should
take large $n$, so we can approximate the spherical harmonics as
\eq{
	P_{2n}(\cos\theta)\approx \sqrt{\frac{4}{\pi}}\cos(2n\theta).
	\label{1.6}
}
The first order correction to the energy in  perturbation theory now yields
\eq{
	\delta E^2=\int\limits_0^{\pi/2}d\theta\,
	\Psi_{2n}^\star(0,\theta)\,m^2\lambda \sin^2\theta\,
	\Psi_{2n}(\theta)
	=\frac{m^2 \lambda}{2}.
	\label{1.7}
}
Finally, the corrected energy levels yield
\begin{equation}
E = \sqrt {(2n + 4)^2 + \frac{{{m^2}\lambda }}{2}} .
\end{equation}
Therefore, one can calculate the anomalous dimension of the corresponding YM operators\footnote{See \cite{min} for more
	details.}
\begin{equation}
(\Delta-4)^2=4n^2+\delta E^2,
\end{equation}
or up to first order in $\lambda$:
\eq{
	\Delta  - 4 = 2n\left( {1 + \frac{1}{2}{\mkern 1mu} \frac{{{m^2}\lambda }}{{{{(2n)}^2}}}} \right).
	\label{1.8}
}
On should note that in this case the $R$-charge is zero. One can include it by considering a pulsating string on $S^5$, which center
of mass is moving on the $S^3$ subspace of $S^5$ \cite{minzar}. While, in the
previous example, $S^3$ part of the metric was assumed trivial, now we
consider all the $S^3$ angles to depend on $\tau$ (only). The
corresponding Nambu-Goto action is then given by
\eq{
	S=-m\sqrt{\lambda}\int\,
	dt\sin\theta\,\sqrt{1-\dot\theta^2-\cos^2\theta
		g_{ij}\dot\phi^i\dot\phi^j },
	\label{1.9}
}
where $\phi_i$ are $S^3$ angles and $g_{ij}$ is the corresponding
$S^3$ metric. In this case the Hamiltonian is written by \cite{minzar}
\eq{
	H=\sqrt{\Pi_\theta^2+\frac{g^{ij}\Pi_i\Pi_j}{\cos^2\theta}
		+m^2\lambda\sin^2\theta},
	\label{1.10}
}
where again the squared Hamiltonian looks like the point
particle one. However, now the potential has angular
dependence. Denoting the quantum number of $S^3$ and $S^5$ by $J$ and
$L$ correspondingly, one can write the Schr\"odinger equation in the form
\eq{
	-\frac{4}{\omega}\dfrac{d}{d\omega}\Psi(\omega)
	+\frac{J(J+1)}{\omega}\Psi(\omega) = L(L+4) \Psi(\omega),
	\label{1.11}
}
with $\omega=\cos^2\theta$. The explicit solution is
\eq{
	\Psi(\omega)=\frac{\sqrt{2(l+1)}}{(l-j)!}\,\frac{1}{\omega}\left(
	\frac{d}{d\omega}\right)^{l-j}
	\omega^{l+j}(1-\omega)^{l-j}
	, \quad j=\frac{J}{2},\quad l=\frac{L}{2}.
	\label{1.12}
}
The first order correction to the squared energy $\delta E^2$ in this case yields

\eq{
	\delta E^2={m^2\lambda\,\frac{2(l+1)^2-(j+1)^2-j^2}{(2l+1)(2l+3)}}.
	\label{1.13}
}
Now, the corrected energy (up to first order in $\lambda$) can be written in the form
	\eq{
		E
	 = \sqrt {L(L + 4)}  + \frac{{ {m^2}\lambda (L - J)(J + L)}}{{4{L^2}\sqrt {L(L + 4)} }}+\mathcal{O}(\lambda^2).
		\label{1.14}
	}
Finally, the anomalous dimension can also be calculated
\eq{
	\gamma=\frac{m^2\lambda}{4L}\alpha(2-\alpha),
	\label{1.15}
}
with $\alpha=1-J/L$.
%

\subsection{Pulsating string solutions in Schr\"odinger background}

Now, we are in a position to obtain pulsating string solutions in Schr\"odinger background. To this end we will construct the Polyakov action and find appropriate solutions. To start with, we write the line element of  Schr\"odinger geometry in global coordinates
\begin{equation}
\frac{ds^2_{Schr_5}}{\ell^2}=-\left(1+\frac{\hat{\mu}^2}{Z^4}+\frac{\vec{X}^2}{Z^2}\right)dT^2+\frac{2dTdV+d\vec{X}^2+dZ^2}{Z^2},
\end{equation}
and
\begin{align} 
\frac{ds^2_{S^5}}{\ell^2}&=\,d\chi^2+d\mu^2+\frac{1}{4}\sin^2\mu\left(d\alpha^2+d\theta^2+d\phi^2\right)\nonumber\\
&+\sin^2\mu\,d\chi\,d\alpha+\sin^2\mu\cos\theta\,d\chi\,d\phi+\frac{\sin^2\mu\cos\theta}{2}\,d\alpha\,d\phi,
\end{align}
where we have redefine $\hat{\chi}\to \chi$ for simplicity, as opposed to Eq. (\ref{eq_S5_metric}).   Due to the TsT transformations the theory acquires also a $B$-field  
\begin{equation}
B_{(2)} =
\dfrac{\ell^2\hat{\mu}}{\alpha^\prime Z^2} dT\wedge\left( d\chi + \dfrac{\sin^{2} \mu}{2} d\alpha +\dfrac{\sin^{2} \mu \cos  \theta}{2} d\phi \right) \,\, .
\end{equation}	
Hence, the Polyakov string action in conformal gauge	($\alpha,\beta=0,1$ and $M,N=0,\dots, 9$) is given by
\begin{equation}
S=-\frac{1}{4 \pi \alpha ^ \prime }\int d\tau d\sigma \left\{\sqrt{-h}\,h^{\alpha \beta} \partial_\alpha X^M \partial_\beta X^N G_{MN}-\epsilon^{\alpha\beta} \partial_\alpha X^M \partial_\beta X^NB_{MN}\right\},
\end{equation}
where $h^{\alpha\beta}=\rm{diag}(-1,1)$ and $\epsilon^{01}=-\epsilon^{10}=1$. The explicit form of the Lagrangian, with respect to the metric and the $B$-field, yields
\begin{align}
-4\pi\alpha'\mathcal{L}&=\,G_{TT}(T'^2-\dot{T}^2)+G_{\vec{X}\vec{X}}(\vec{X}'^2-\dot{\vec{X}}^2)+G_{ZZ}(Z'^2-\dot{Z}^2)\nonumber\\
&+2G_{TV}(T'V'-\dot{T}\dot{V})+G_{\chi\chi}(\chi'^2-\dot{\chi}^2)+G_{\mu\mu}(\mu'^2-\dot{\mu}^2)\nonumber\\
&+G_{\theta\theta}(\theta'^2-\dot{\theta}^2)+G_{\alpha\alpha}(\alpha'^2-\dot{\alpha}^2)+G_{\phi\phi}(\phi'^2-\dot{\phi}^2)\nonumber\\
&+2G_{\chi\alpha}(\chi'\alpha'-\dot{\chi}\dot{\alpha})+2G_{\chi\phi}(\chi'\phi'-\dot{\chi}\dot{\phi})+2G_{\alpha\phi}(\alpha'\phi'-\dot{\alpha}\dot{\phi})\nonumber\\
&+2B_{T\chi}(T'\dot{\chi}-\dot{T}\chi')+2B_{T\alpha}(T'\dot{\alpha}-\dot{T}\alpha')+2B_{T\phi}(T'\dot{\phi}-\dot{T}\phi').
\end{align}
Beside the equations of motion (EoM) the solutions should also satisfy the Virasoro constraints
\begin{align}
&\text{Vir}_1:\qquad G_{MN}\left(\dot{X}^M\dot{X}^N+X'^MX'^N\right)=0\,\label{vir1},\\
&\text{Vir}_2:\qquad G_{MN}\,\dot{X}^MX'^N=0.\label{vir2}
\end{align}
To find the general solution of the set of EoMs, satisfying \eqref{vir1} and \eqref{vir2}, is quite nontrivial task.
To obtain pulsating string solutions in Schr\"odinger background we need an appropriate ansatz for circular string configurations, 
\al{
&T =\kappa \tau,\,\kappa>0,\quad\, Z=const\neq 0,\quad\, \vec{X}=\vec{0},\quad V=0, \nonumber \\
&\mu=\mu (\tau),\quad\, \theta=\theta (\tau),\quad \chi =n^1 \sigma,\quad
\alpha=n^2 \sigma,\quad\, \phi =n^3 \sigma.\label{ansatz}
}
A quick check shows that the second Virasoro constraint is trivially satisfied, whereas the first one gives raise to the following nontrivial equation:
\begin{align}
\dot{\mu}^2+\frac{1}{4}\sin^2\mu\left(\dot{\theta}^2+(n^2)^2+(n^3)^2+4n^1(n^2+n^3\cos\theta)+2n^2n^3\cos\theta\right)\nonumber\\
-\left(1+\frac{\hat{\mu}^2}{Z^4}\right)\kappa^2+(n^1)^2=0.  
\end{align}
The substitution of the ansatz \eqref{ansatz} into the equations for $V,\,\vec{X},\,\chi,\,\alpha$ and $\phi$, shows that they are trivially satisfied. The other equations, namely for $T$, $Z$, $\theta$ and $\mu$, require further analysis. Let us list the relevant equations. The EoM for $T$ results in
\begin{equation}\label{PulsCond}
\frac{d}{d\tau}\left[\sin^2\mu\left(n^2+n^3\cos\theta\right)\right]=0\quad\Rightarrow\quad \sin^2\mu\left(n^2+n^3 \cos\theta\right)=A=const.
\end{equation}
We refer to this equation as the pulsating condition. For $Z$ we have the solution
\begin{equation}
\label{Z:}
Z^2=\frac{2\hat{\mu}\,\alpha'\kappa}{2n^1+A},
\end{equation}
which shows that $Z$ is a constant, as it should be, due to the ansatz (\ref{ansatz}).
The equation for $\theta$ becomes
\begin{equation}
\frac{d}{d\tau}\left(\sin^2\mu\,\dot{\theta}\right)\,-\,n^3 \sin^2\mu\,\sin\theta\left(2n^1+n^2-\frac{2\hat{\mu}\kappa}{\alpha'Z^2}\right)=0.
\end{equation}
Finally, for $\mu$ we have
\begin{equation}
\ddot \mu + \frac{1}{4}\sin \mu \cos \mu \left((n^2)^2 + (n^3)^2 - \dot \theta^2 + 2n^2n^3 \cos \theta + 4\left(n^1-\frac{\hat\mu \kappa}{\alpha'Z^2} \right)(n^2+n^3\cos \theta )\right) = 0.
\end{equation}
Obviously $n^1$ and $A$ can not be zero at the same time, due to \eqref{Z:}, thus $2n^1+A\neq 0$. This suggests the following cases:
\begin{align}
(i) & \qquad n^1\neq 0,\quad A\neq 0 ,\\
(ii) & \qquad n^1 =0,\quad A\neq 0 ,\\
(iii) & \qquad n^1\neq 0,\quad A=0 ,\quad \text{such as} \quad n^2+n^3\cos\theta =0, \quad \text{or} \quad n^2=n^3 =0.
\end{align}
Let us focus on the first case $(i)$. The substitution of the pulsating condition \eqref{PulsCond}, 
\begin{equation}
\label{PulsCondition}
\sin^2 \mu =\frac{A}{n^2+n^3\cos\theta},
\end{equation}
into the $\theta$ equation, results in the following relation
\begin{equation}
\label{3.35}
\frac{d}{d\tau}\left(\dfrac{\dot{\theta}}{n^2+n^3\cos\theta}\right)\,-\,n^3 \,K\,\frac{\sin\theta}{n^2+n^3\cos\theta}=0,
\end{equation}
where $K=2n^1+n^2-\frac{2\,\hat{\mu}\,\kappa}{\alpha'Z^2}$	.
Multiplying Eq. \eqref{3.35}  by $\dfrac{\dot{\theta}}{n^2+n^3\cos\theta}$, one finds
\begin{equation}
\frac{d}{d\tau}\left[\dfrac{1}{2}\left( \dfrac{\dot{\theta}}{n^2+n^3\cos\theta}\right)^2 \right]\,-\,\frac{d}{d\tau} \left( \dfrac{K}{n^2+n^3\cos\theta} \right)=0. 
\end{equation}	
Integrating once we end up with an equation for the only variable	needed for a complete pulsating string solution:
\begin{equation}
\dfrac{1}{2}\left( \dfrac{\dot{\theta}}{n^2+n^3\cos\theta}\right)^2 \,-\, \dfrac{K}{n^2+n^3\cos\theta} =C=const.
\end{equation}
We can rewrite the latter equation such as
\begin{equation}
\label{3.38}
\dot{\theta} ^2= 2 (n^2+n^3\cos\theta)\left(C\,(n^2+n^3\cos\theta)+K \right). 
\end{equation}
Multiplying Eq. \eqref{3.38} by $\sin^2 \theta$, we find
\begin{equation}
\label{3.39}
\left(\frac{d}{d\tau} \cos\theta\right)^2=2(1-\cos^2\theta) (n^2+n^3\cos\theta)(Cn^3\cos\theta+Cn^2 +K).
\end{equation} 
At this point, it is convenient to introduce a new variable $u=\cos\theta, |u| \leq 1$. Thus, equation \eqref{3.39} becomes
\begin{equation}
\left(\frac{d}{d\tau}\, u \right)^2=2\,(1-u^2\,)\,P_2\,(u),
\label{eq-u}
\end{equation}
where 
\begin{equation}
P_2\,(u)=C(n^3)^2 \, u^2 + n^3\,(2Cn^2 +K)\,u+n^2\,(Cn^2 +K).
\end{equation}
Due to \eqref{eq-u}, the following condition should hold $P_2\,(u)\geq 0, \,\,\forall \, u\,\in\,[-1,\,1]$. Since the discriminant of the polynomial $P_2(u)$ is non-negative, $D=K^2\, (n^3)^2\geq 0$,  there are two real roots\footnote{Note that the integers $n^i$ could be positive or negative. On the other hand we have the constraint $-1\leq u\leq 1$ since $u=\cos\theta$. If one of the roots is outside of this interval, due to \eqref{eq-u}, there are no real solutions to this equation. The case when both roots are outside the interval is discussed below.}:
\eq{
u_{1,2}=\frac{1}{2Cn^3}\left(-2Cn^2-K \pm K \right)=
\begin{cases}
-\frac{n^2}{n^3}, \\
-\left(\frac{n^2}{n^3}+\frac{K}{Cn^3}\right).
\end{cases}
\label{roots}
}
Note that the first case coincides with the case $K=0$. 	
Now, equation \eqref{eq-u} takes the form
\begin{equation}\label{EquationU}
\left(\dfrac{d}{d\tau} \,u \right)^2=2\,C\left( n^3\right) ^2\,(1-u^2\,)\,(u-u_1)\,(u-u_2)\geq 0.
\end{equation}
The range of $u_1$ and $u_2$ can be arbitrary. It is clear that the solutions essentially depend on the signs of $C,\,u_1,\,u_2$ and that of $u$ between $[-1,1]$. Therefore, the problem of finding periodic solutions is brought to several cases, which should be analyzed separately. 
For example, for $C>0$ and $K>0$, the following orders are possible:
\begin{subequations}
\begin{align}
\begin{split}
u_1<u_2<-1\leq u(\tau) \leq 1,
\end{split}\\
\begin{split}
u_1<-1<u_2 \leq u(\tau) \leq 1,
\end{split}\\
\begin{split}
-1< u_1<u_2\leq u(\tau)\leq 1,
\end{split}\\
\begin{split}
-1\leq u(\tau) \leq u_1<u_2<1,
\end{split}\\
\begin{split}
-1\leq u(\tau) \leq u_1<1<u_2,
\end{split}\\
\begin{split}
-1\leq u(\tau) \leq 1<u_1<u_2.
\end{split}
\end{align}
\end{subequations}
In the case of $C<0$ and $K>0$, one has the following options:
\begin{subequations}
\begin{align}
\begin{split}
u_1<-1\leq u(\tau) \leq u_2<1,
\end{split}\\
\begin{split}
u_1<-1\leq u(\tau)\leq 1<u_2,
\end{split}\\
\begin{split}
-1<u_1\leq u(\tau) \leq u_2 <1,
\end{split}\\
\begin{split}
-1< u_1\leq u(\tau) \leq 1<u_2.
\end{split}
\end{align}
\end{subequations}
Finally, it is also possible that $K=0$, and $C>0$. In this case
\begin{equation}
P_2\,(u)=C\left(n^3\right)^2\,u^2 + 2Cn^2n^3\,u+C\left(n^2\right)^2\geq 0
\end{equation}
The discriminant is zero, thus $u_1=u_2=-n^2/n^3$ and Eq. \eqref{eq-u} takes the form
\begin{equation}
\left(\dfrac{d}{d\tau} \,u \right)^2=2\,C\left( n^3\right) ^2\,(1-u^2\,)\,(u-u_1)^2\geq 0.
\end{equation}
We start our analysis with $C>0,\,\,K>0$ and  $u_1<u_2<-1\leq u(\tau)\leq 1$. Then, we integrate the equation \eqref{EquationU}:
\begin{equation}
\int\limits_{u(\tau)}^1 \,\frac{du}{\sqrt{(1-u)(u-(-1))(u-u_2)(u-u_1)}} \,=\, \sqrt{2C\,(n^3)^2}\, \int\limits_{0}^{\tau}\,d\tau\,=\, \sqrt{2C\,(n^3)^2}\,\tau.
\label{integral-1}
\end{equation}
This integral can be found in  Gradshteyn and Ryzhik \cite{GrandshteynR}, which is an integral of the type
\begin{equation}
\int\limits_{b}^{u} \,\frac{dz}{\sqrt{(a-z)(z-b)(z-c)(z-d)}}, \quad a\geq u \geq b > c > d .
\end{equation}
This integral can be solved in terms of a first kind elliptic integral:
\begin{equation}\label{Elliptic}
\int\limits_{b}^{u} \,\frac{dz}{\sqrt{(a-z)(z-b)(z-c)(z-d)}} \,=\, \dfrac{2}{\sqrt{(a-c)(b-d)}} \, F(\xi,\, r),
\end{equation}
where 
\begin{equation}
\xi=\arcsin \sqrt{\dfrac{(b-d)(a-u)}{(a-b)(u-d)}},\qquad r=\sqrt{\dfrac{(a-b)(c-d)}{(a-c)(b-d)}}.
\end{equation}
Therefore, in view of \eqref{integral-1}, we have
\begin{equation}
F(\xi,\, r)=\, \sqrt{2^{-1}C\,(n^3)^2 (a-c)(b-d)}\,\tau,
\end{equation}
where $a=1,\,b=-1,\, c=u_2 =-|u_2|,\, d=u_1 =-|u_1|$.
In other words, one finds
\begin{equation}
\sin \xi = \sn\left(\sqrt{2^{-1}C\,(n^3)^2 (a-c)(b-d)}\,\tau ,\,r\right),
\end{equation}
or explicitly
\begin{equation}
\dfrac{(b-d)(a-u)}{(a-b)(u-d)}=\, \sn^2\left( \sqrt{2^{-1}C\,(n^3)^2 (a-c)(b-d)}\,\tau ,\, r\right).
\end{equation}
The solution of this equation for $u(\tau)$ is given by
\begin{equation}
u(\tau)=\frac{1-\dfrac{2|u_1|}{|u_1|-1}\, \sn^2 \left( \sqrt{2^{-1}C\,(n^3)^2 (|u_1|-1)(1+|u_2|)}\,\tau ,\, r\right)}{1+\dfrac{2}{|u_1|-1}\, \sn^2 \left( \sqrt{2^{-1}C\,(n^3)^2 (|u_1|-1)(1+|u_2|)}\,\tau ,\, r\right)}\,.
\end{equation}
Returning to the original variable the final solution yields
\begin{equation}
\theta (\tau)=\arccos\left( \frac{1-\dfrac{2|u_1|}{|u_1|-1}\, \sn^2 \left( \sqrt{2^{-1}C\,(n^3)^2 (|u_1|-1)(1+|u_2|)}\,\tau ,\, r\right)}{1+\dfrac{2}{|u_1|-1}\, \sn^2 \left(\sqrt{2^{-1}C\,(n^3)^2 (|u_1|-1)(1+|u_2|)}\,\tau ,\, r\right)}\right). 
\end{equation} 
We list the solutions for the other cases from $(i),(ii)$ and $(iii)$ in appendix \ref{appendix_B}.
%

\section{Energy corrections and anomalous dimensions}

To proceed with the energy corrections we need the Nambu-Goto string action.
The first step towards finding the Nambu-Goto action is to make a pullback of the line element of the metric of  $Schr_5 \times S^5$ to the subspace, where string dynamics takes place. The result for the metric is
\begin{equation}\label{pulsMetric}
ds^2 =\ell^2 \left(  -\left|G_{00} \right| dT^2+\sum\limits_{i,j=1} ^2 G_{ij}(\mu,\theta)d x^i d x^j + \sum\limits_{k,h=1} ^3 \hat{G}_{kh}(\mu,\theta)d y^k d y^h  \right), 
\end{equation}
where for brevity we used the notations
\al{ 
& \mu=\mu(\tau), \qquad \theta(\tau)=x^2(\tau), \qquad \chi=y^1,\qquad \alpha=y^2,\qquad \phi=y^3, \nonumber \\	
& \left|G_{00} \right|= 1+ \dfrac{\hat{\mu}^2}{Z^4},\qquad
	\left( G_{ij} \right) =
	\left(\begin{matrix}
		1 & 0 \\
		0 & \dfrac{\sin^{2} \mu}{4}
	\end{matrix}\right),\label{G00a}
}
and 
\eq{
	\left(\hat{G}_{kh}\right)=\left(\begin{matrix}
		1 & \dfrac{\sin^{2} \mu}{2}& \dfrac{\sin^{2} \mu \cos  \theta}{2} \\
		\dfrac{\sin^{2} \mu}{2} & \dfrac{\sin^{2} \mu}{4}& \dfrac{\sin^{2} \mu \cos  \theta}{4}\\
		\dfrac{\sin^{2} \mu \cos  \theta}{2}&\dfrac{\sin^{2} \mu \cos  \theta}{4} &  \dfrac{\sin^{2} \mu}{4}
	\end{matrix}\right).
}
In contrast to the case from the previous subsection, the TsT transformation generates a $B$-field, involving time direction:
\begin{equation}\label{B-field}
B_{(2)} =\ell^2 \dfrac{\hat{\mu}}{\alpha^\prime Z^2} dT\wedge\left( d\chi + \dfrac{\sin^{2} \mu}{2} d\alpha +\dfrac{\sin^{2} \mu \cos  \theta}{2} d\phi \right)= \ell^2\sum\limits_{k=1} ^3 b_{0k} (\mu,\theta)\,dT\wedge dy^k,
\end{equation}
where
\eq{ 
	b_{01}=\dfrac{\hat{\mu}}{\alpha^\prime Z^2}\,,  \qquad   b_{02}=\dfrac{\hat{\mu}}{2\alpha^\prime Z^2}\sin^{2}\mu\,,  \qquad   b_{03}=\dfrac{\hat{\mu}}{2\alpha^\prime Z^2}\sin^{2}\mu\,\cos\theta.
	\label{notation-hamiltonian}
}
The induced worldsheet metric and the $B$-field components become
\begin{subequations}\label{ws}
	\begin{align}
	\frac{ds^2 _{ws}}{\ell^2}&= \left(  -\left|G_{00} \right| \kappa^2+\sum\limits_{i,j=1} ^2 G_{ij}(\mu,\theta)\dot{x}^i \dot{x}^j\right)d\tau^2  + \left( \sum\limits_{k,h=1} ^3 \hat{G}_{kh}(\mu,\theta) n^k n^h \right) d\sigma^2 , \label{ws-m}\\
	B_{(2)}^{ws}&=2\ell^2 B_{\tau\sigma}\,d\tau\wedge d\sigma ,\qquad B_{\tau\sigma}(\mu,\theta)=-B_{\sigma\tau}=\dfrac{1}{2}\sum\limits_{k=1} ^3 b_{0k} (\mu,\theta)\,\kappa\,n^k  \label{ws-b}.
	\end{align}
\end{subequations}
With these preparations at hand we can write the Nambu-Goto action, 
\begin{equation} \label{NG}
S_{NG}= -\dfrac{1}{2 \pi \alpha ^ \prime } \int d\tau d\sigma \sqrt{ - \det \left(G_{\mu\nu}\,\partial_\alpha X^{\mu}\,\partial_\beta X^{\nu} - B_{\mu\nu}\,\partial_\alpha X^{\mu}\,\partial_\beta X^{\nu}   \right) }\,,
\end{equation}
or in terms of the notations we introduced above:
\begin{equation} \label{NG-ws}
S_{NG}^{puls}= -\dfrac{\ell^2}{\alpha^\prime }\int d\tau \sqrt{ \left( \sum\limits_{k,h=1} ^3 \hat{G}_{kh}(\mu,\theta)\,n^k n^h \right) \left( \left|G_{00} \right| \kappa^2  - \sum\limits_{i,j=1} ^2 G_{ij}(\mu,\theta)\dot{x}^i \dot{x}^j\right)  - B_{\tau\sigma} ^2 (\mu,\theta) }\,,
\end{equation}
where $\dfrac{\ell^2}{\alpha ^ \prime }=\sqrt{\lambda}$\, is the 't Hooft coupling constant.

\indent To obtain the energy correction we consider the Hamiltonian formulation of the problem. For the case under consideration the canonical momenta are given by
\begin{equation}\label{momenta}
\Pi_k =\frac{\partial L}{\partial \dot{x}^k}= \dfrac{\sqrt{\lambda}\left( \sum\limits_{k,h=1} ^3 \hat{G}_{kh}\,n^k n^h \right) \sum\limits_{i=1} ^2 G_{ki}\,\dot{x}^i } {\sqrt{ \left( \sum\limits_{k,h=1} ^3 \hat{G}_{kh}\,n^k n^h \right) \left( \left|G_{00} \right| \kappa^2  - \sum\limits_{i,j=1} ^2 G_{ij}\,\dot{x}^i \dot{x}^j\right) - B_{\tau\sigma} ^2 } }\,,\quad k=1,2 .
\end{equation}
Solving for the derivatives in terms of the canonical momenta and substituting back into the Legendre transformed Lagrangian, $L=\Pi_k\,\dot{x}^k-H$, we find the (square of) the Hamiltonian:
\begin{align}\label{Hamiltonian1}
H_{puls}^2 = \frac{{||\vec n|{|^2}\left| {{G_{00}}} \right|{\kappa ^2} - B_{\tau \sigma }^2}}{{||\vec n|{|^2}}}\sum\limits_{i,j = 1}^2 {{G^{ij}}} {\Pi _i}{\Pi _j} + \lambda \left( {||\vec n|{|^2}\left| {{G_{00}}} \right|{\kappa ^2} - B_{\tau \sigma }^2} \right),
\end{align}
where ${||\vec n|{|^2}}$ is defined in Eq. (\ref{A2}). 
As in other cases of pulsating strings in holography, we observe that $H^2$ looks like a point-particle Hamiltonian, in which the last term serves as a potential
\begin{equation}\label{potential}
U(\mu,\theta)=\left(\sum\limits_{k,h=1} ^3 \hat{G}_{kh}(\mu,\theta)\,n^k n^h \right)  \left|G_{00} \right| \kappa^2 - B_{\tau\sigma}^2 (\mu,\theta).
\end{equation}
The steps we will follow involve a perturbative expansion in the small coupling constant $\lambda$. The semiclassical quantization of the pulsating string then ends up with the corrections to the energy. According to the AdS/CFT dictionary, the anomalous dimension of the corresponding SYM operators are directly related to the corrections to the energy.

 For brevity of the notations it is useful to make some definitions. First, we define the following ''scalar product'':
\eq{
\left\| \vec{n}\right\|^2:=\sum\limits_{k,h=1} ^3 \hat{G}_{kh}\,n^k n^h =\dfrac{1}{4} \left(a_1 +a_2 \sin^2 \mu \right)>0,
\label{A2}
}
where the constants $a_1$ and $a_2$ are written by
\eq{
a_1=4(n^1)^2+2(2n^1+n^2)A\,, \qquad a_2= (n^3)^2-(n^2)^2.
\label{A2-a}
}
Here, $A=\left( n^2+n^3 \cos\theta\right) \sin ^2 \mu=const$, is the familiar pulsating condition from Eq. (\ref{PulsCond}). Analogously, we find
\begin{equation}\label{A3}
B_{\tau\sigma} ^2 = \dfrac{\kappa^2b^2}{4}, \qquad\quad b^2= \dfrac{1}{4} \dfrac{\hat{\mu}^2}{{\alpha^\prime}^2 Z^4}\left( A+ 2 n^1\right) ^2=\frac{{{{(A + 2{n^1})}^4}}}{{4{{\alpha'}^4}{\kappa ^2}}},
\end{equation}
where $A$ is the defining constant in the pulsating condition (\ref{PulsCond}), $\hat\mu$ is defined in Eq. (\ref{B-global-12}), and $Z$ is defined in Eq. (\ref{Z:}).
Hence, one has
\begin{equation}
\left|G_{00} \right| \kappa^2 \left\| \vec{n}\right\|^2  - B_{\tau\sigma} ^2  = \dfrac{\kappa^2}{4} \left( a_1 \left|G_{00} \right|-b^2 + a_2 \left|G_{00} \right| \sin^2\mu\right)=\dfrac{\kappa^2}{4}\left( A_1 + A_2 \sin^2 \mu \right), 
\end{equation}
with
\begin{equation}
A_1=a_1 \left|G_{00} \right|-b^2,\qquad\quad A_2=a_2 \left|G_{00} \right|
\end{equation}
The kinetic term of the Hamiltonian \eqref{Hamiltonian1} is considered as a two dimensional Laplace-Beltrami operator
\begin{equation}
\sum\limits_{i,j=1} ^2 G^{ij}(\mu,\theta)\,\Pi_i\Pi_j \,\,\longrightarrow \,\, \triangle^{(2)} =\frac{1}{\sqrt{\det (G_{ij})}} \partial_i \left( \sqrt{\det (G_{ij})}\,\, G^{ij}\, \partial_j\right).
\end{equation}
In our case, this operator takes the form
\begin{equation}
\triangle^{(2)}=\dfrac{1}{\sin\mu}\dfrac{\partial}{\partial\mu}\left(\sin\mu \dfrac{\partial}{\partial\mu}  \right) +\dfrac{4}{\sin^2 \mu} \dfrac{\partial^2}{\partial\theta ^2} .
\end{equation}
This Laplace-Beltrami operator defines the eigen-functions of the Hamiltonian. Collecting the results and using the above notations the Schr\"odinger equation for the wave function becomes
\begin{equation}
\dfrac{\left|G_{00} \right| \kappa^2 \left\| \vec{n}\right\|^2  - B_{\tau\sigma} ^2}{\left\| \vec{n}\right\|^2} \,\,\triangle^{(2)}\,\,\Psi(\mu,\theta)= -E^2 \,\,\Psi(\mu,\theta),
\end{equation}
or explicitly
\begin{equation}\label{Shrodinger}
\left( \dfrac{1}{\sin\mu}\dfrac{\partial}{\partial\mu}\left(\sin\mu \dfrac{\partial}{\partial\mu}  \right) +\dfrac{4}{\sin^2 \mu} \dfrac{\partial^2}{\partial\theta ^2} \right) \Psi(\mu,\theta)=-\,\dfrac{E^2\left(a_1 +a_2 \sin^2 \mu \right)}{\kappa ^2\left( A_1 + A_2 \sin^2 \mu \right)} \, \Psi(\mu,\theta).
\end{equation}
Now, we make the standard separation of the variables
\begin{equation}\label{eqLquantumNumber}
\Psi(\mu, \theta)=\varPsi(\mu)\,e^{il\theta},\qquad\quad l \in \mathbb{Z}.
\end{equation}
The equation for $\varPsi(\mu)$ yields
\begin{equation}\label{mu}
\left( \dfrac{1}{\sin\mu}\dfrac{d}{d\mu}\left(\sin\mu \dfrac{d}{d\mu}  \right) -\dfrac{4l^2}{\sin^2 \mu}  \right) \varPsi(\mu)=-\,\dfrac{E^2\left(a_1 +a_2 \sin^2 \mu \right)}{\kappa ^2\left( A_1 + A_2 \sin^2 \mu \right)} \, \varPsi(\mu).
\end{equation}
To bring Eq. \eqref{mu}, into more familiar from, we introduce a new variable $z=\cos\mu$, where $0\leq z\leq 1$ and $0 \leq\mu \leq \pi/2$. Thus, one can rewrite the above equation as 
\begin{equation}\label{EQH}
\left( \left( 1-z^2 \right) \dfrac{d^2}{dz^2}\,-2z \dfrac{d}{dz}\,-\,  
\dfrac{4l^2}{1-z^2}\, + \,\dfrac{E^2\left(a_1 +a_2 \left(1-z^2 \right)\right)}{\kappa ^2\left( a_1 \left|G_{00} \right|-b^2 + a_2\left|G_{00} \right| \left(1-z^2 \right) \right)}  \right) \varPsi(z)=0.
\end{equation}
The latter is a second order ordinary differential equation of the form
\begin{equation}\label{eqTheFullEquation}
 \varPsi''(z) + \tilde P(z)\, \varPsi '(z) + \tilde Q(z)\, \varPsi(z) = 0,
\end{equation}
where
\begin{equation}\label{eqOriginalHeun}
\tilde P(z) =-\,\frac{2z}{1 - z^2},\qquad \tilde Q(z) =-\,\frac{1}{1 - z^2}\left( \frac{4l^2}{1 - z^2}\,-\,\frac{E^2\,\,\left(a_1 + a_2\left(1 - z^2\right)\right)}{\kappa ^2|G_{00}|\left(a_1 + a_2(1 - z^2) - \frac{b^2}{|G_{00}|}\right)} \right).
\end{equation}
In general, this is a Heun type equation,
\begin{equation}
H''(z) + P(z)\,H'(z) + Q(z)\,H(z) = 0\,,
\end{equation}
with coefficients
\begin{equation}
P(z) = \left( \frac{\gamma }{z} + \frac{\delta }{z - 1} + \frac{\varepsilon }{z - a} \right),\qquad Q(z) = \frac{\alpha \beta z - q}{z(z - 1)(z - a)},
\end{equation}
where $\varepsilon=\alpha+\beta+1-\gamma-\delta$. Now, we want to map our equation to Heun's one. In order to find the corresponding parameters $q,a,\alpha,\beta,\gamma,\delta$, we make the identification $P(z)=\tilde{P}(z)$ and $Q(z)=\tilde{Q}(z)$. After equating the coefficients in front of the powers of $z$, one finds the following algebraic system:
\begin{equation}
\left| \begin{array}{l}
a\gamma = 0,\\
1-\alpha-\beta= 0,\\
{a_2}|{G_{00}}|{\kappa ^2}q = 0,\\
a(\gamma  + \delta  - 2) - \delta  = 0,\\
(a - 1)\delta  + \alpha  + \beta  + 1 = 0,\\
\left( E^2 + \alpha \beta |G_{00}|\kappa ^2 \right)a_2 = 0,\\
\left( b^2 - (a_1 + a_2)|G_{00}|\right)\kappa ^2q = 0,\\
\left( b^2 - (a_1 + 2a_2)|G_{00}| \right)\kappa ^2q = 0,\\
4a_2|G_{00}|\kappa ^2l^2 - (a_1 + 2a_2)E^2 + \left( b^2 - (a_1 + 2a_2)|G_{00}|\right)\kappa ^2\alpha \beta  = 0,\\
\left( (a_1 + a_2)|G_{00}|-4b^2 \right)\kappa^2l^2-(a_2 + a_1)E^2 + \left(b^2 - (a_1 + 2a_2)|G_{00}|\right)\kappa ^2 \alpha \beta  = 0.
\end{array} \right.
\end{equation}
This system has a solution only if the following set of constraints, $\{ |G_{00}| > 0,\,\kappa>0,\,E>0,\,b>0,\,a_1>0,\,a_2\ne 0,\,l \in \mathbb{Z}\}$, is obeyed. The explicit solutions for the parameters are given by
\begin{equation}
q = 0,\;\;\;a =  - 1,\;\;\;\gamma  = 0,\;\;\;\delta  = 1,\;\;\;\alpha  = \frac{1}{2} \pm \frac{\sqrt {16|G_{00}|l^2a_2+b^2} }{2b},\;\;\;\beta  = 1 - \alpha ,
\end{equation}
accompanied with two more constraints:
\begin{equation}\label{a_12}
a_1=\frac{b^2}{|G_{00}|},\qquad a_2=\frac{b^2E^2}{4|G_{00}|^2\kappa^2 l^2}\,.
\end{equation}
Therefore, Eq. (\ref{eqOriginalHeun}) becomes the Legendre equation
\begin{equation}\label{eqLegendre}
\varPsi''(z)+\left(\frac{1}{z-1}+\frac{1}{z+1} \right)\varPsi '(z) + \frac{\alpha (1-\alpha )}{(z-1)(z+1)}\,\varPsi(z)=0.
\end{equation}
Quantization condition tells us that the solutions are the Legendre $P$ and $Q$ polynomials:
\begin{equation}
\varPsi(z)=c_1P_{\alpha-1}(z)+c_2\,Q_{\alpha-1}(z).
\label{solution-f}
\end{equation}
On the other hand, the wave function must be square integrable with respect to the measure  
\begin{equation}
\sqrt{\det(G_{ij})}\,d\mu\,d\theta=\dfrac{\sin\mu}{2}\,d\mu\,d\theta=-\dfrac{1}{2}\,dz\,d\theta,\qquad 0\leq\theta\leq\pi, \quad 0 \leq z\leq 1.
\end{equation}
The requirement of quadratic integrability sets $c_2=0$, so the solution to Eq. \eqref{solution-f} becomes $\varPsi(z)=cP_n(z)$. Therefore, the normalized wave functions (\ref{Shrodinger}) have the following explicit form
\begin{equation}\label{normalizedwave}
\Psi_{n,\,l}(z,\theta)=\sqrt{\frac{2(2n+1)}{\pi}}\,\,P_n (z)\, e^{il\theta},\qquad \quad l, n \in \mathbb{Z}.
\end{equation}
Let us turn to the energy spectrum. For $\varPsi(z)=P_{\alpha-1=n}(z)$ one can easily find
\begin{equation}\label{spectrum}
E^2 = n(n+1)|G_{00}|\kappa ^2=n\left( {n + 1} \right)\kappa\left( {{\kappa } + \frac{{{b }}}{{2}}} \right),
\end{equation}
where the constant $b$ is defined in Eq. (\ref{A3}).
Taking the square root one arrives at the energy spectrum
\begin{equation}\label{eqEroot}
E = \sqrt {n\left( {n + 1} \right)\kappa \left( {\kappa  + \frac{b}{2}} \right)} .
\end{equation}

We can now proceed with corrections to the energy. To do that, it is convenient to write the potential (\ref{potential}) in the form
\begin{equation}
U(\mu)=\frac{\kappa^2}{4}\,\left(\left|G_{00} \right|(a_1+a_2\sin^2\mu)-b^2\right).
\end{equation}
It is more convenient to work in terms of the new variable $z$, where the potential becomes
\begin{equation}\label{potential_z}
U(z)=\frac{\kappa^2}{4}\,\left( \left|G_{00} \right| \left(a_1 +a_2 (1-z^2)\right)- b^2\right).
\end{equation}
However, it is easy to observe that due to the additional constraints (\ref{a_12}) and (\ref{spectrum}) the expression for the potential greatly simplifies:
\begin{equation}\label{potential_z1}
U(z)=\frac{\kappa^2 b^2 n(n+1)}{16\,l^2}\, (1-z^2).
\end{equation}
We should note that in the last expression we used the original constants from the string ansatz.
In the perturbation theory first correction to the energy is given then by the expression:
\begin{equation}
\delta E ^2=\frac{\lambda}{2} \int\limits_{z=0}^{z=1} \int\limits_{\theta=0}^{\theta=\pi} \left| \Psi_{n,\,l}(z,\theta) \right| ^2 \, U(z)\,dz\,d\theta=\lambda\,\frac{\kappa^2 b^2 n(n+1)(2n+1)}{16\,l^2}\int\limits_{0}^{1}(1-z^2)\left| P_n (z)\right|^2dz.
\end{equation}
Hence, for the first quantum correction to the energy, we find
\begin{align*}
\delta E ^2\,=& \lambda\,\frac{\kappa^2 b^2 n(n+1)(2n+1)}{16\,l^2}\,\int\limits_{0}^{1}(1-z^2)\left| P_n (z)\right|^2 dz\nonumber\\
=&\lambda\,\frac{\kappa^2 b^2 n(n+1)}{16\,l^2}\left((2n+1)\int\limits_{0}^{1} P_n^2(z)\,dz-(2n+1)\int\limits_{0}^{1} z^2 P_n^2(z)\,dz\right) \nonumber \\
=&\lambda\,\frac{\kappa^2 b^2 n(n+1)}{16\,l^2}\,\left(1-\frac{2n^2+2n-1}{(2n-1)(2n+3)}\right)=\lambda\,\frac{\kappa^2 b^2 }{8\,l^2}\,\frac{n(n+1)(n^2+n-1)}{(2n-1)(2n+3)},
\end{align*}
thus one has
\begin{equation}
\delta E^2=\lambda\,\frac{\kappa^2 b^2 }{8\,l^2}\,\frac{n(n+1)(n^2+n-1)}{(2n-1)(2n+3)} \,.
\label{1-correction-final-1}
\end{equation}
In order to obtain Eq. (\ref{1-correction-final-1}) we took advantage of the following relations:
\begin{equation}
\int\limits_{0}^{1}P_n ^2(z)\,dz=\dfrac{1}{2n+1}, \qquad \int\limits_{0}^{1} z^2 P_n^2 (z)\,dz=\frac{2n^2 +2n-1}{(2n-1)(2n+1)(2n+3)}.
\end{equation}
Although we will consider only the first correction term, it is straightforward to calculate also higher corrections to the energy, which have the form
\eq{
	\delta E ^2_{(s)}\propto \int\limits_{0}^{1}\, (1-z^2)^s\left| P_n (z)\right|^2\,dz=
	\sum_{k=0}^s (-1)^k\begin{pmatrix} s \\ k \end{pmatrix} \int\limits_{0}^{1} x^{2k} \left| P_n (z)\right|^2\,dz.
}
The last integral is evaluated in \cite{Carlitz:1961}:
\begin{equation}
\int\limits_{0}^{1} x^{2k} \left| P_n (z)\right|^2\,dz=2^{-2n}\frac{(2n)!}{(n!)^2}\frac{\Gamma(2k+1)\,{}_3F_2\left(-n,-n,\frac{2(k-n)+1}{2},-2n,\,k-n+1\right)}{\left(\dfrac{2(k-n)+3}{2}\right)_{2n}\Gamma(2(k-n+1))}.
\end{equation}

Combining Eqs. (\ref{spectrum}) and Eq. (\ref{1-correction-final-1}) one finds the  corrected energy
\begin{equation}\label{spectrumCorr}
E =\sqrt {n\left( {n + 1} \right)\kappa } {\left( {\kappa  + \frac{b}{2} + {\mkern 1mu} \frac{{\kappa {b^2}}}{{8{\mkern 1mu} {l^2}}}{\mkern 1mu} \frac{{({n^2} + n - 1)}}{{(2n - 1)(2n + 3)}}\lambda } \right)^{1/2}}\,.
\end{equation}
By expanding the square root up to first order in $\lambda$,
\begin{equation}
E=\frac{1}{2}\sqrt {2\kappa n(n + 1)(b + 2\kappa )} \left( {1 + \frac{{{b^2}{\kappa }\lambda \left( {{n^2} + n - 1} \right)}}{{8{l^2}(4n(n + 1) - 3)(b + 2\kappa )}}} \right)+\mathcal{O}(\lambda^2),
\end{equation}
one can calculate the anomalous dimension
\begin{equation}
\Delta  = \frac{{{b^2}{\kappa}\lambda \left( {{n^2} + n - 1} \right)\sqrt {2\kappa n(n + 1)(b + 2\kappa )} }}{{16{l^2}(4n(n + 1) - 3)(b + 2\kappa )}}.
\end{equation}

Considering $L\neq 0$, from the first twisted boundary condition Eq. (\ref{eq_boundary_condition_1}) one finds that the string only pulsates in $Schr_5$, i.e. $J=0$. This is due to the fact that $x^-$ does not depend on $\sigma$, as it is obvious from Eqs. (\ref{eq_A2}) and (\ref{ansatz}).  On the other hand, the second boundary condition (\ref{eq_boundary_condition_2}) leads to the following relation between the winding number $n^1$ and the shift $L$:
\begin{equation}\label{eq_n1_L_relation}
n^1=m-\frac{LP_{-}}{2\pi},
\end{equation}
where we have used the ansatz for $\chi=n^1\sigma$. From equation (\ref{eq_n1_L_relation}) we find that the energy (\ref{spectrum}) and its first quantum correction (\ref{1-correction-final-1}) explicitly depend on the shift $L$.  
 \\
\indent Finally, the classical energy of the string is given by
\begin{equation}
E_{cl}^{} = -\int_{ - L/2}^{L/2} {d\sigma \frac{{\partial {\cal L}}}{{\partial ({\partial _\tau }T(\tau ,\sigma ))}}}  =- \frac{{L{\ell ^2}}}{{4{{\alpha '}^2}\pi }}\left( {\frac{{\hat \mu }}{{{Z^2}}}(2{n^1} + A) - 2\alpha '\kappa \left( {\frac{{{{\hat \mu }^2}}}{{{Z^4}}} + 1} \right)} \right)=\frac{L\ell^2\kappa}{2\pi\alpha'}\,,
\end{equation}
where
\begin{equation}
\frac{{\hat \mu }}{{{Z^2}}} = \frac{{2{n^1} + A}}{{2\alpha '\kappa }}.
\end{equation}
One notes that the classical energy depends explicitly on the shift $L$. 

At the end of this section we summarize our results for the energy spectrum of the pulsating strings in $Schr_5\times S^5$ and the anomalous dimensions of the operators in the corresponding dual gauge theory. The uncorrected energy levels $E$ of the quantized system of pulsating strings are given by
\begin{equation}\label{}
E = \sqrt {n\left( {n + 1} \right)\kappa \left( {\kappa  + \frac{b}{2}} \right)} ,
\end{equation}
where $n$ is the principal quantum number and $b$ is the contribution from the $B$-field in Eq. (\ref{A3}).
The 1-loop corrected energy $E_{\text{(1-loop)}}$ yields
\begin{equation}\label{eqOneLoopCorrectedE}
E_{\text{(1-loop)}} =\sqrt {n\left( {n + 1} \right)\kappa } {\left( {\kappa  + \frac{b}{2} + {\mkern 1mu} \frac{{\kappa {b^2}}}{{8{\mkern 1mu} {l^2}}}{\mkern 1mu} \frac{{({n^2} + n - 1)}}{{(2n - 1)(2n + 3)}}\lambda } \right)^{1/2}}\,,
\end{equation}
where $\lambda$ is the t' Hooft coupling constant and $l$ is an angular quantum number from Eq. {\ref{eqLquantumNumber}}.
The anomalous dimension $\Delta$ of the Young-Mills operators from the dual gauge theory reads
\begin{equation}
\Delta  = \frac{{{b^2}{\kappa}\lambda \left( {{n^2} + n - 1} \right)\sqrt {2\kappa n(n + 1)(b + 2\kappa )} }}{{16{l^2}(4n(n + 1) - 3)(b + 2\kappa )}}.
\end{equation}
Finally, the corrected energy levels up to arbitrary orders in perturbation theory can be written by
\begin{align}
\nonumber E_{total}&=\sqrt{E^2+\sum_{s=1}^{\infty}\delta E ^2_{(s)}}= 
\left[ {n\left( {n + 1} \right)\kappa \left( {\kappa  + \frac{b}{2}} \right) + \lambda {\mkern 1mu} \frac{{{\kappa ^2}{b^2}n(n + 1)(2n + 1){2^{ - 2n}}(2n)!}}{{16{\mkern 1mu} {l^2}{{(n!)}^2}}}} \right.\\
&{\left. { \times \sum\limits_{s = 1}^\infty  {\sum\limits_{k = 0}^s {{{( - 1)}^k}} \left( {\begin{array}{*{20}{c}}
				s\\
				k
				\end{array}} \right)\frac{{\Gamma (2k + 1){{\mkern 1mu} _3}{F_2}\left( { - n, - n,\frac{{2(k - n) + 1}}{2}, - 2n,{\mkern 1mu} k - n + 1} \right)}}{{{{\left( {\frac{{2(k - n) + 3}}{2}} \right)}_{2n}}\Gamma (2(k - n + 1))}}} } \right]^{1/2}}\,.
\end{align}

It is important to mention that the exact one-loop correction (\ref{eqOneLoopCorrectedE}) in the considered Schr\"{o}dinger background may contain also contribution from the fermionic degrees of freedom. However, finding this contribution is fairly complicated even in the case of unbroken supersymmetry \cite{Beccaria:2010zn}. Therefore, we leave this subtle question for a future work.
\section{Conclusions}

In this paper, we have studied a class of pulsating string solutions in a Schr\"odinger background. We provided a brief review of the deformations generating the
Schr\"odinger geometry. As a particular case of Drinfel'd-Reshetikhin twist, the solution generating TsT deformations  involve the time direction and one spatial dimension in the internal space. The deformation is known also as null Melvin twist transformation.

The study of pulsating strings in particular backgrounds with external fluxes has been implemented for instance in \cite{Banerjee:2015bia}  and appeared to be nontrivial. From this perspective we expected the study of our problem to be also nontrivial.   
To find pulsating string solutions in Schrödinger background, we have adopted
an appropriate pulsating ansatz for the circular string configuration.
Considering the bosonic part of the string action in conformal gauge, we
have found the relevant equations of motion and Virasoro constraints. The
problem of finding periodic solutions imposes certain conditions even after the choice of pulsating string ansatz. The restriction of the parameters leads to several non-trivial
cases, all of which have analytic solutions as combinations of trigonometric
and Jacobi elliptic functions.

An important key point of pulsating strings is that they reduce the problem to the (squared)
Hamiltonian, which has the form of a point-particle Hamiltonian. It allows clearly to distinguish  an effective string potential, which, being multiplied by the coupling $\lambda$,  is used later for perturbative expansion.  Obtaining the effective wave function for the problem, we used perturbation theory to obtain the
corrections to the energy. The latter, due to AdS/CFT correspondence, are
related to the anomalous dimensions of operators in the dual gauge theory. The
unperturbed energy spectrum turned out to be equidistant and independent on
the components of the $B$-field. On the other hand, the first correction picked up a factor proportional to the $B$-field. Furthermore, we have also
obtained the explicit expressions for the higher order corrections in terms
of generalized hypergeometric functions. 

To the best of our knowledge, the obtained solutions are the first pulsating type solutions in Schr\"odinger backgrounds and in backgrounds with non-relativistic gauge dual in general. Thus, the energy and the obtained corrections are the first examples of contributions from this string sector to the anomalous dimensions in non-relativistic dual gauge theory. One should note that giant magnon solutions in spherical part extended in Schr\"odinger space have been found in \cite{Georgiou:2017pvi} (see also \cite{Ahn:2017bio}) and all the results are complementary to each other. 

There are several interesting questions, which could be addressed further.
First of all, it would be interesting to identify the operators on the gauge theory side whose anomalous dimension correspond to  the various
contributions in the energy spectrum. Some operators have been discussed in \cite{Guica:2017jmq}. We didn't touch these issues, but it is worth to note that the wave functions obtained in two different approaches in \cite{Guica:2017jmq} are also Legendre polynomials as in our considerations. The operators discussed in \cite{Guica:2017jmq} could give a clue in this direction.  Next, one can consider perturbations
around the pulsating string solutions, which, at the quantum level, could
give contributions to the scaling relations. It is also of great
interest to perform an analysis comparing our results to that in gauge
theory side, such as finding the Yang-Mills operators corresponding to these
string configurations. The study of these problems in some other backgrounds, $Schr_5\times T^{1,1}$ for instance, could also be of particular interest. Finally, one should consider that the exact one-loop correction (\ref{eqOneLoopCorrectedE}) contains potential contributions from the fermionic sector.
However, this contribution is fairly
complicated even in the case of unbroken supersymmetry \cite{Beccaria:2010zn}.  In the case of
 Schr\"{o}dinger backgrounds, fermionic contribution  is even more subtle
question and remains to be done. We are planning to  address these issue in
a future publication.

\paragraph{Acknowledgements}\ \\
R. R. is grateful to Kostya Zarembo for discussions on various issues of holography in Schr\"odinger backgrounds. T. V. and M. R are grateful to Prof. G.~Gjordjevic for the warm hospitality at the University of Ni\v{s}, where some of these results have been presented. This work was supported in part by BNSF Grant DN-18/1 and H-28/5.

\begin{appendix}

\section{A remark on more general deformations}
\label{appA}

As we mentioned in the main text, the most general deformations preserving integrability containing the TsT techniques, is based on the so-called Drinfe’ld-Reshetikhin (DR) twist. In this Appendix we present a brief review of the DR twist following mainly \cite{Guica:2017jmq}.

\indent To this end, let us consider the scattering matrix $\mathbb{S}$. The DR twist of the scattering matrix is realized as
\eq{
	\mathbb{S}\:\rightarrow\: \tilde{\mathbb{S}}=\mathbb{F}\,\mathbb{S}\,\mathbb{F},\qquad \mathbb{F}=e^{\frac{i}{2}\sum_{i,j}\gamma_{ij}(H_i\otimes H_j-H_j\otimes H_i)},
}
where $H_i$ are Cartan elements of the isometry group, $\gamma_{ij}$ is a constant antisymmetric matrix. In the cases above, instead of coefficient times Cartan element, we have the corresponding charges. It is important that all the charges are commuting (they could be also supercharges). For instance, for the case of Cartan elements we have
\eq{
	(H_i)_{kj}=\frac{1}{2}\left(\delta_{i,k}\delta_{i,j}- \delta_{i+1,j}\delta_{i+1,k}\right).
}

The most relevant case is to consider the R-matrix, which acts on the tensor product of two vector spaces:
\eq{
	R_{ab}(u): \quad V_a\otimes V_b\:\longrightarrow	\: V_a\otimes V_b,
}
and satisfy the Yang-Baxter (YB) equation
\eq{
	R_{ab}(u-v)R_{ac}(u)R_{bc}(v)=R_{bc}(v)R_{ac}(u)R_{ab}(u-v).
}
Then Drinfe'ld twist is the most general linear transformation of the form
\eq{
	R_{ab}(u)\:\longrightarrow\: \tilde{R}_{ab}(u)= F_{ab}R_{ab}(u)F_{ab}, 
}
which preserves integrability. The constant matrix $F_{ab}$ satisfies the following conditions:
\begin{itemize}
\item It is a constant solution of YB equations:
\eq{F_{ab}F_{ac}F_{bc}=F_{bc}F_{ac}F_{ab}.}
\item Obeys associativity condition: constant solution of YB equation with an intertwining relation with the R-matrix:
\eq{
		R_{ab}(u)F_{ca}F_{cb}=F_{cb}F_{ca}R_{ab}(u).}
\item To preserve regularity of the R-matrix, the twist should satisfy an additional constraint dubbed to as unitarity condition:
\eq{
		F_{ab}F_{ba}=1.}
\end{itemize}
Looking at the discussion in the beginning of this Section, one can see that a wide class of solutions is associated with the  commuting charges. For instance, having a set of commuting in $V_a$ charges $Q^i_a$, $[Q^i_a,Q^j_a]=0$, the condition 
\eq{
	e^{i\omega_k Q^k_a}e^{i\omega_l Q^l_b}R_{ab}(u)= R_{ab}(u)e^{i\omega_k Q^k_a}e^{i\omega_l Q^l_b}
}
for each $k,l$ means that $e^{i\omega_k Q^k_a}$ is a non-degenerate linear transformation on $V_a$ and a symmetry of $R_{ab}$. It is simple exercise to check that the following $F_{ab}$ satisfies YB equation:
\eq{
	F_{ab}=K_a K^{-1}_b, \qquad K_a=e^{i\omega_i Q^i_a}.
}
%
%
%
To summarize, one can construct the Drinfe'ld twist operator as (summation over $i,j$ is understood):
\eq{
		F_{ab}=e^{\frac{i}{2}\gamma_{ij}Q^i_aQ^j_b},
}
where $\gamma_{ij}=-\gamma_{ji}$.

\section{Schr\"odinger space in Global coordinates}
We start with the observation that the Schr\"odinger metric \eqref{schro-metric} in local coordinates  consists of two pieces
\eq{
	ds^2=-\ell^2 \frac{{\hat\mu}^2(dx^+)^2}{z^4} + ds^2_{AdS_5},
	\label{schro-metric-a}
}
where the second part is the $AdS_5$ metric in  light-cone coordinates 
\begin{equation}
\label{A.1}
ds^2_{AdS_5}=\frac{\ell^2}{z^2}\left(2dx^+dx^-+d\vec{x}^2+dz\right).
\end{equation}
We first focus on the AdS piece.
To obtain the metric in global coordinates we make the following transformations
\begin{equation}\label{eq_A2}
x^+=\tan T,\quad x^-=V-\frac{1}{2}\left(Z^2+\vec{X}^2\right)\tan T,\quad z=\frac{Z}{\cos T},\quad \vec{x}=\frac{\vec{X}}{\cos T}.
\end{equation}
Thus 
\begin{align}
\label{A.3}
&dx^+=\frac{dT}{\cos^2T},\qquad dx^-=dV-\tan T\left(ZdZ+\vec{X}.d\vec{X}\right)-\frac{1}{2}\left(Z^2+\vec{X}^2\right)\frac{dT}{\cos^2T},\nonumber\\
&dz=\frac{dZ}{\cos T}+\frac{Z\sin TdT}{\cos^2T},\qquad d\vec{x}=\frac{1}{\cos T}\left(d\vec{X}+\vec{X}\tan TdT\right).
\end{align}
Substituting \eqref{A.3} into \eqref{A.1} we find the metric of $AdS_5$ in global coordinates
\begin{equation}
ds^2_{AdS_5}=\frac{\ell^2}{Z^2}\left(2dTdV-(Z^2+\vec{X}^2)dT^2+d\vec{X}^2+dZ^2\right).
\end{equation}
To obtain the first part we just need the relation 
\begin{equation}
\frac{\hat{\mu}^2}{z^4}{dx^+}^2=\frac{\hat{\mu}^2}{Z^4}{dT}^2.
\end{equation}
Putting everything together we obtain the Schr\"odinger metric and the $B-$field in global coordinates in the form
\begin{equation}
\frac{ds^2_{Schr_5}}{\ell^2}=-\left(\frac{\hat{\mu}^2}{Z^4}+1 \right)dT^2+
\frac{2dT\,dV-\vec{X}^2dT^2+d\vec{X}^2+dZ^2}{Z^2},
\label{metric-schro-global}
\end{equation}
\begin{equation}
\alpha' B_{(2)}= \frac{\ell^2\hat{\mu}\, dT}{Z^2}\wedge (d\hat{\chi}+P).
\label{B-global-1}
\end{equation}

\section{Other pulsating string solutions}\label{appendix_B}
In this appendix we consider the solutions for the other cases from $(i),\,(ii)$ and $(iii)$.

\begin{itemize}
\item For $C>0, K>0$ and $u_1<-1<u_2\leq u(\tau)\leq 1$:
\begin{equation}
\theta (\tau)=\arccos\left( \frac{1-\dfrac{1-u_2}{|u_1|+u_2}|u_1|\, \sn^2 \left(\sqrt{C\,(n^3)^2\,(|u_1|+u_2)}\,\tau ,\, r\right)}{1+\dfrac{1-u_2}{|u_1|+u_2}\, \sn^2 \left(\sqrt{C\,(n^3)^2\,(|u_1|+u_2)}\,\tau ,\, r\right)}\right). 
\end{equation}
\item For $C>0, K>0$ and $-1 < u_1 < u_2 \leq u(\tau) \leq 1$:
\begin{equation}
\theta (\tau)=\arccos\left( \frac{1-\dfrac{1-u_2}{1+u_2}\, \sn^2 \left(\sqrt{2^{-1}C\,(n^3)^2\,(1-u_1)(1+u_2)}\,\tau ,\, r\right)}{1+\dfrac{1-u_2}{1+u_2}\, \sn^2 \left(\sqrt{2^{-1}C\,(n^3)^2\,(1-u_1)(1+u_2)}\,\tau ,\, r\right)}\right). 
\end{equation}
\item For $C>0, K>0$ and $-1 \leq u(\tau) \leq u_1<u_2<1$:
\begin{equation}
\theta (\tau)=\arccos\left( \frac{u_1-\dfrac{1+u_1}{1+u_2}u_2\,\, \sn^2 \left(\sqrt{2^{-1}C\,(n^3)^2\,(1-u_1)(1+u_2)}\,\tau ,\, r\right)}{1-\dfrac{1+u_1}{1+u_2}\, \sn^2 \left(\sqrt{2^{-1}C\,(n^3)^2\,(1-u_1)(1+u_2)}\,\tau ,\, r\right)}\right). 
\end{equation}
\item For $C>0, K>0$ and $-1 \leq u(\tau) \leq 1<u_1<u_2$:
\begin{equation}
\theta (\tau)=\arccos\left( \frac{1-\dfrac{2u_1}{1+u_1}\, \sn^2 \left(\sqrt{2^{-1}\,(n^3)^2\,(1+u_1)(u_2-1)}\,\tau ,\, r\right)}{1-\dfrac{2}{1+u_1}\, \sn^2 \left(\sqrt{2^{-1}C\,(n^3)^2\,(1+u_1)(u_2-1)}\,\tau ,\, r\right)}\right).
\end{equation}
\item For $C>0, K>0$ and $-1 \leq u(\tau) \leq u_1<1<u_2$:
\begin{equation}
\theta (\tau)=\arccos\left( \frac{u_1-\dfrac{1+u_1}{2}\, \sn^2 \left(\sqrt{C\,(n^3)^2\,(u_2-u_1)}\,\tau ,\, r\right)}{1-\dfrac{1+u_1}{2}\, \sn^2 \left(\sqrt{C\,(n^3)^2\,(u_2-u_1)}\,\tau ,\, r\right)}\right).
\end{equation}
\item For $C<0,\,K>0,\,\,u_1<-1 \leq u(\tau) \leq u_2<1$\,\, and \,\,\,$q=\sqrt{\dfrac{(b-c)(a-d)}{(a-c)(b-d)}}$:
\begin{equation}
\theta (\tau)=\arccos\left( \frac{u_2-\dfrac{1+u_2}{2}\, \sn^2 \left(\sqrt{|C|\,(n^3)^2\,(u_2+|u_1|)}\,\tau ,\, q\right)}{1-\dfrac{1+u_2}{2}\, \sn^2 \left(\sqrt{|C|\,(n^3)^2\,(u_2+|u_1|)}\,\tau ,\, q\right)}\right).
\end{equation}
\item For $C<0,\,K>0$ and $u_1<-1 \leq u(\tau) \leq 1<u_2$:
\begin{equation}
\theta (\tau)=\arccos\left( \frac{1-\dfrac{2u_2}{1+u_2}\,\sn^2\left(\sqrt{2^{-1}|C|\,(n^3)^2\,(1+|u_1|)(1+u_2)}\,\tau ,\, q\right)}{1-\dfrac{2}{1+u_2}\, \sn^2 \left(\sqrt{2^{-1}|C|\,(n^3)^2\,(1+|u_1|)(1+u_2)}\,\tau ,\, q\right)}\right).
\end{equation}
\item For $C<0,\,K>0$ and $-1<u_1 \leq u(\tau) \leq u_2<1$:
\begin{equation}
\theta (\tau)=\arccos\left( \frac{u_2-\dfrac{u_2-u_1}{1-u_1}\, \sn^2 \left(\sqrt{2^{-1}|C|\,(n^3)^2\,(1-u_1)(1+u_2)}\,\tau ,\, q\right)}{1-\dfrac{u_2-u_1}{1-u_1}\, \sn^2 \left(\sqrt{2^{-1}|C|\,(n^3)^2\,(1-u_1)(1+u_2)}\,\tau ,\, q\right)}\right).
\end{equation}
\item For $C<0,\,K>0$ and $-1<u_1\leq u(\tau) \leq 1<u_2$:
\begin{equation}
\theta (\tau)=\arccos\left( \frac{1-\dfrac{1-u_1}{u_2-u_1}u_2\, \sn^2 \left(\sqrt{|C|\,(n^3)^2\,(u_2-u_1)}\,\tau ,\, q\right)}{1-\dfrac{1-u_1}{u_2-u_1}\, \sn^2 \,\left(\sqrt{|C|\,(n^3)^2\,(u_2-u_1)}\,\tau ,\, q\right)}\right).
\end{equation}
\end{itemize}
For $C>0$ and $K=0$, we have
\begin{equation}
\int\limits_{1}^{u} \,\frac{du}{|u-u_1|\sqrt{1-u^2}}=-\sqrt{2C(n^3)^2}\int\limits_{0}^{\tau}d\tau.
\end{equation}
Hence,
\begin{equation}
\int\limits_{0}^{\theta} \,\frac{d\theta}{|\cos\theta-u_1|}=\sqrt{2C(n^3)^2}\,\tau.
\end{equation}
There are four possible cases:
\begin{subequations}
\begin{align}
\begin{split}
u_1<-1 \leq u(\tau)\leq 1,
\end{split}\\
\begin{split}
-1<u_1\leq u(\tau)\leq 1,
\end{split}\\
\begin{split}
-1\leq u(\tau)<u_1<1,
\end{split}\\
\begin{split}
-1\leq u(\tau)\leq 1<u_1.
\end{split}
\end{align}
\end{subequations}
The solutions are:
\begin{itemize}
\item For $C>0, K=0$ and $u_1<-1 \leq u(\tau)\leq 1$:
\begin{equation}
\tan\left( \frac{\theta}{2} \right)=\sqrt{\frac{|u_1|+1}{|u_1|-1}}\,\tan\left(\sqrt{2^{-1}C(n^3)^2(u_1^2-1)}\,\tau\right).
\end{equation}
Using the formula $\cos\theta=\dfrac{1-\tan^2\frac{\theta}{2}\,}{1+\tan^2\frac{\theta}{2}\,}$, we evaluate
\begin{equation}
\theta(\tau)=\arccos\left(\frac{1-\dfrac{|u_1|+1}{|u_1|-1}\,\tan^2\left(\sqrt{2^{-1}C(n^3)^2(u_1^2-1)}\,\tau\right)}{1+\dfrac{|u_1|+1}{|u_1|-1}\,\tan^2\left(\sqrt{2^{-1}C(n^3)^2(u_1^2-1)}\,\tau\right)}\right).
\end{equation}
\item For $C>0, K=0$ and $-1\leq u(\tau)\leq 1<u_1$:
\begin{equation}
\tan\left(\frac{\theta}{2}\right)= \sqrt{\frac{u_1-1}{u_1+1}}\,\tan\left(\sqrt{2^{-1}C(n^3)^2(u_1^2-1)}\,\tau\right),
\end{equation}
or
\begin{equation}
\theta(\tau)= \arccos\left(\frac{1-\dfrac{u_1-1}{u_1+1}\,\tan^2\left(\sqrt{2^{-1}C(n^3)^2(u_1^2-1)}\,\tau\right)}{1+\dfrac{u_1-1}{u_1+1}\,\tan^2\left(\sqrt{2^{-1}C(n^3)^2(u_1^2-1)}\,\tau\right)}\right).
\end{equation}
\item In the following cases $(C>0, K=0)$: $-1<u_1\leq u(\tau)\leq 1$ and $-1\leq u(\tau)<u_1<1$ we do not have periodic solutions. 
\end{itemize}

\indent Now we consider the case $(ii)$, where $n^1=0,\,A\neq0$. The first Virasoro constraint provides
\begin{eqnarray}
\dot{\mu}^2+\frac{1}{4}\sin^2\mu\left(\dot{\theta}^2+(n^2)^2+(n^3)^2+2n^2n^3\cos\theta\right)-\left(1+\frac{\hat{\mu}^2}{Z^4}\right)\kappa^2=0.  
\end{eqnarray}
The other equations are as follows.
\begin{itemize}
\item  For Z:
\begin{equation}
Z^2=\frac{2\hat{\mu}\alpha'\kappa}{A}.
\end{equation}
\item The next equation is for $\theta$:
\begin{equation}
\frac{d}{d\tau}\left(\sin^2\mu\,\dot{\theta}\right)\,-\,n^3 \sin^2\mu\,\sin\theta\left(n^2-\frac{2\hat{\mu}\kappa}{\alpha'Z^2}\right)=0.
\end{equation}
\item Finally, for $\mu$ one has
\begin{equation}
\ddot{\mu}+\frac{1}{4}\sin\mu\cos\mu\,\left((n^2)^2+(n^3)^2-\dot{\theta}^2+2n^2 n^3 \cos\theta-\frac{4\hat{\mu}\kappa}{\alpha'Z^2}(n^2+n^3\cos\theta)\right)=0.
\end{equation}
\end{itemize}
Once again we substitute the pulsating condition (\ref{PulsCondition}) into the equation for $\theta$ to obtain an ordinary differential equation with respect only to $\theta$:
\begin{equation}
\frac{d}{d\tau}\left(\dfrac{\dot{\theta}}{n^2+n^3\cos\theta}\right)\,-\,n^3K'\,\frac{\sin\theta}{n^2+n^3\cos\theta}=0,
\end{equation} 
where $K'=n^2-\frac{2\hat{\mu}\kappa}{\alpha'Z^2}$. Thus, the case $(ii)$, is the same as $(i)$, but with different constant $K'$.

\indent The last case we consider is $(iii):$ $ n^1\neq0,\,A=0$. Setting $A=0$, we have to solve
\begin{equation}
\label{A.6}
A=\sin^2\mu\left(n^2+n^3 \cos\theta\right)=0.
\end{equation}
So, the first option is
\begin{equation}
n^2+n^3\cos\theta=0\quad\Rightarrow\quad \cos\theta=-\frac{n^2}{n^3}\quad\Rightarrow\quad |n^2|\leq|n^3|,\,\,\,\theta=const.
\end{equation}
Next step is to check the first Virasoro constraint:
\begin{equation}
\label{A.8}
\dot{\mu}^2=\left(1+\frac{\hat{\mu}^2}{Z^4}\right)\kappa^2-(n^1)^2-\frac{1}{4}\sin^2\mu\left((n^3)^2-(n^2)^2\right).
\end{equation}
The equations of motion are as follows:
\begin{itemize}
\item For $Z$ we find
\begin{equation}
\label{A.26}
Z^2=\frac{\hat{\mu}\alpha'\kappa}{n^1}.
\end{equation}
\item The equation of motion for $\theta$ provides the relation
\begin{equation}
\label{A.27}
Z^2=\frac{2\hat{\mu}\kappa}{\alpha'(2n^1+n^2)}.
\end{equation}
\item The only non-constant equation is for $\mu$:
\begin{equation}
\label{A.11}
\dot{\mu}^2=N^2-\frac{1}{4}\sin^2\mu\left((n^3)^2-(n^2)^2\right).
\end{equation}
\end{itemize}
From Eqs. \eqref{A.26} and \eqref{A.27} we find the following relation between the windings $n^1$ and $n^2$:  
\begin{equation}
\frac{n^2}{n^1}=2\left(\frac{1}{\alpha'^2}-1\right).
\end{equation}
Using Eqs. \eqref{A.8}, \eqref{A.26} and \eqref{A.11}, one can express for the constant $N^2$ such as
\begin{equation}
N^2=\kappa^2+(n^1)^2\left(\frac{1}{\alpha'^2}-1\right).
\end{equation}
Now, we are in  a position to integrate the equation for $\mu$:
\begin{equation}
\int\limits_{0}^{\mu}\frac{d\mu}{\sqrt{1-k^2\sin^2\mu}}=F\left(\mu,k\right)=\pm|N|\int\limits_{0}^{\tau}d\tau,
\end{equation}
where $4N^2k^2=(n^3)^2-(n^2)^2$. Finally, it is straightforward to obtain 
\eq{ 
\sin\mu=\pm\,\sn(|N|\tau,k),
}
or explicitly for $\mu$  
\begin{equation}
\mu(\tau)=\pm\arcsin\left(\sn(|N|\tau,k)\right).
\end{equation}
\indent The second solution of Eq. (\ref{A.6}) is $n^2=n^3=0$.  The corresponding non-zero Virasoro constraint becomes
\begin{equation}
\label{A.16}
\dot{\mu}^2+\frac{1}{4}\sin^2\mu\,\,\dot{\theta}^2-N^2=0.
\end{equation}
The equations of motion provide the following relations:
\begin{itemize}
\item For the constant $Z$ one obtains
\begin{equation}
Z^2=\frac{\hat{\mu}\alpha'\kappa}{n^1}.
\end{equation}
\item  The equation for $\theta$ gives
\begin{equation}
\label{A.18}
\frac{d}{d\tau}\left(\sin^2\mu\,\dot{\theta}\right)=0.
\end{equation}
\item  The equation for $\mu$ is also non-trivial:
\begin{equation}
\label{A.19}
\frac{d}{d\tau}\left(\dot{\mu}\right)-\frac{1}{4}\sin\mu\cos\mu\,\dot{\theta}^2=0.
\end{equation}
\end{itemize}
Now we multiply Eq. (\ref{A.19}) by $\sin\mu$ and use Eq.(\ref{A.16}) to express $\sin^2\mu\,\dot{\theta}^2=-4(\dot{\mu}^2-N^2)$. The resulting 	equation is given by
\begin{equation}
\sin\mu\frac{d}{d\tau}(\dot{\mu})+\cos\mu\,(\dot{\mu}^2-N^2)=0.
\end{equation}
By the fact that $\frac{d}{d\tau}(\dot{\mu}\sin\mu)=\sin\mu\,\frac{d}{d\tau}(\dot{\mu})+\dot{\mu}^2\cos\mu$, the last equation becomes
\begin{equation}
\frac{d}{d\tau}\left(\dot{\mu}\sin\mu\right)-N^2\cos\mu=0,
\end{equation}
or
\begin{equation}
\frac{d^2}{d\tau^2}(\cos\mu)+N^2\cos\mu=0.
\end{equation}
Its solution is
\begin{equation}
\label{A.23}
\mu(\tau)=\arccos\left(C_1\cos(N\tau)+C_2\sin(N\tau)\right).
\end{equation}
Integrating Eq. \eqref{A.18} we obtain
\begin{equation}
\sin^2\mu\,\dot{\theta}=D=const.
\end{equation}
We can integrate this equation with respect to $\theta$ and $\tau$:
\begin{equation}
\int\limits_{0}^{\theta} d\theta=\int\limits_{0}^{\tau} \frac{D}{1-\cos^2\mu}d\tau=\int\limits_{0}^{\tau} \frac{D}{1-\left(C_1\cos(N\tau)+C_2\sin(N\tau)\right)^2}\,d\tau,
\end{equation}
where we used Eq. \eqref{A.23}. The solution is
\begin{equation}
\theta(\tau)=\frac{D}{N\sqrt{C_1^2+C_2^2-1}}\, {\rm{arctanh}}\left(\frac{C_1C_2+(C_2^2-1)\,\tan(N\tau)}{\sqrt{C_1^2+C_2^2-1}}\right),
\end{equation}
with an additional constraint $C_1^2+C_2^2<1$ which is a requirement for the solution to be periodic.

\end{appendix}


\end{document}